\documentclass[%
 aip, jcp,
 amsmath,amssymb,
reprint,%
]{revtex4-1}

\usepackage{graphicx}% Include figure files
\usepackage{dcolumn}% Align table columns on decimal point
\usepackage{bm}% bold math
\usepackage[utf8]{inputenc}
\usepackage[T1]{fontenc}
\usepackage{mathptmx}
\usepackage{etoolbox}
\usepackage{bm}
\usepackage{ragged2e}
\usepackage{float}
\usepackage{multirow}
\usepackage{array}
\usepackage[colorlinks = true, linkcolor = blue, urlcolor = blue]{hyperref}

\newcommand{\dirGAPTS}{ACE-Dir}
\newcommand{\indirGAPTS}{ACE-Indir1}
\newcommand{\indirEXP}{ACE-Indir2}

\makeatletter
\def\@email#1#2{%
 \endgroup
 \patchcmd{\titleblock@produce}
  {\frontmatter@RRAPformat}
  {\frontmatter@RRAPformat{\produce@RRAP{*#1\href{mailto:#2}{#2}}}\frontmatter@RRAPformat}
  {}{}
}%
\makeatother
\begin{document}

\preprint{AIP/123-QED}

\title[]{Computationally Efficient Machine-Learned Model for GST \\ Phase Change Materials via Direct and Indirect Learning}
% Force line breaks with \\
\author{Owen R. Dunton}
 \email{odunton@wesleyan.edu}
\affiliation{Department of Physics, Wesleyan University, Middletown, CT 06459, USA}
\author{Tom Arbaugh}
\affiliation{Department of Physics, Wesleyan University, Middletown, CT 06459, USA}
\author{Francis W.\ Starr}
 \email{fstarr@wesleyan.edu}
\affiliation{Department of Physics, Wesleyan University, Middletown, CT 06459, USA}

\date{Submitted 5 November 2024}
%\date{Submitted 26 June 2024}
%\date{\today}% It is always \today, today,
             %  but any date may be explicitly specified

\begin{abstract}
Phase change materials such as Ge$_{2}$Sb$_{2}$Te$_{5}$ (GST) are ideal candidates for next-generation, non-volatile, solid-state memory due to the ability to retain binary data in the amorphous and crystal phases, and rapidly transition between these phases to write/erase information. Thus, there is wide interest in using molecular modeling to study GST. Recently, a Gaussian Approximation Potential (GAP) was trained for GST to reproduce Density Functional Theory (DFT) energies and forces at a fraction of the computational cost [Zhou {\it et al.} Nature Electronics $\mathbf{6}$, 746–754 (2023)]; however, simulations of large length and time scales are still challenging using this GAP model. Here we present a machine-learned (ML) potential for GST implemented using the Atomic Cluster Expansion (ACE) framework.  This ACE potential shows comparable accuracy to the GAP potential but performs orders of magnitude faster. We train the ACE potentials both directly from DFT, as well as using a recently introduced indirect learning approach where the potential is trained instead from an intermediate ML potential, in this case, GAP. Indirect learning allows us to consider a significantly larger training set than could be generated using DFT alone. We compare the directly and indirectly learned potentials and find that both reproduce the structure and thermodynamics predicted by the GAP, and also match experimental measures of GST structure. The speed of the ACE model, particularly when using GPU acceleration, allows us to examine repeated transitions between crystal and amorphous phases in device-scale systems with only modest computational resources.

\end{abstract}

\maketitle

\section{\label{sec:intro} Introduction}

\begin{figure}
\centering
\includegraphics[width=0.48\textwidth,keepaspectratio]{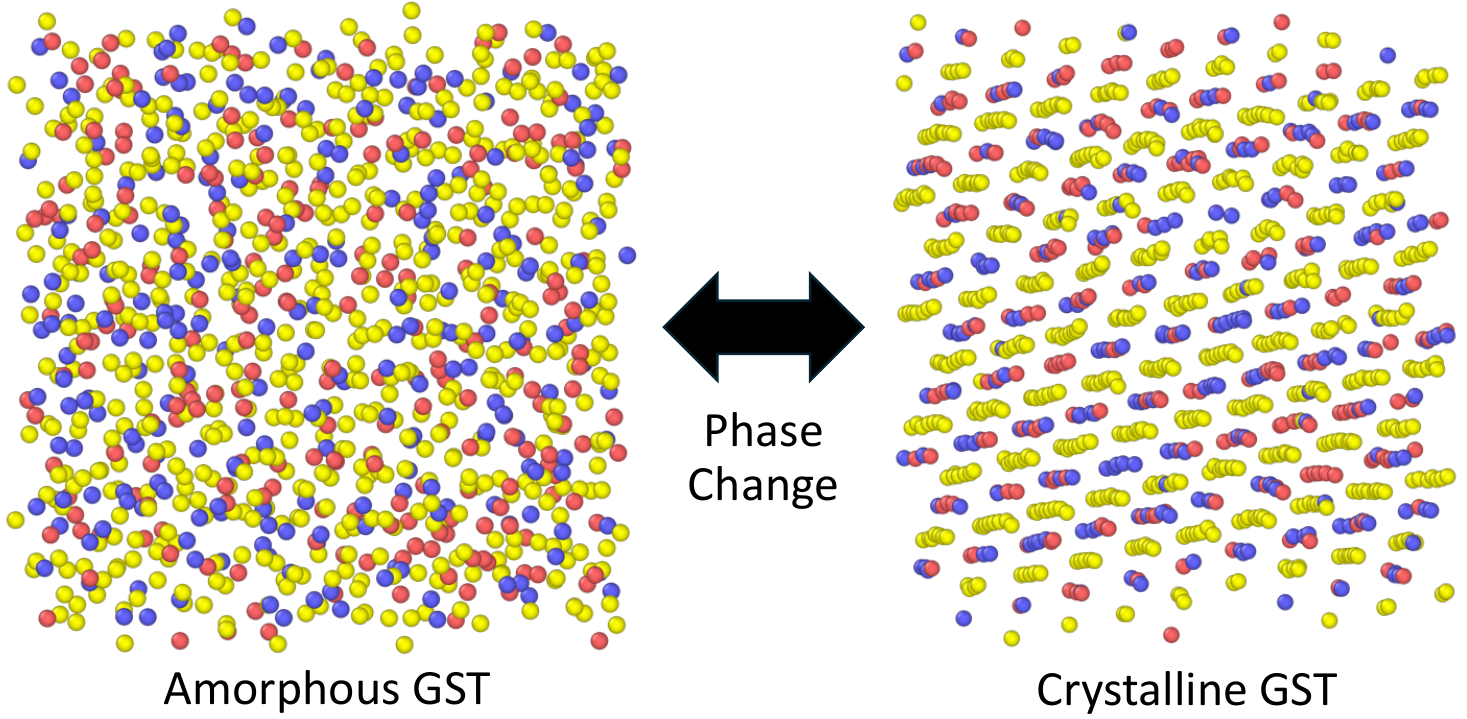}
\caption{A render of our molecular simulations of the amorphous and crystal structures of GST (made using OVITO~\cite{OVITO}).  Ge are shown in red, Sb are shown in blue, and Te are shown in yellow.}
\label{fig:render}
\end{figure}

\justifying

Phase change materials (PCMs) have received attention for their ability to reversibly transition between stable solid phases with different electrical conductivities and optical reflectivities~\cite{ren2011phase, rios2015integrated, ovshinsky1968reversible, lai2003current, le2020overview}. These property differences allow PCMs to encode information in the solid phases -- crystalline and amorphous -- which correspond to bits of binary data~\cite{rios2015integrated, lai2003current, le2020overview}, as illustrated by Fig.~\ref{fig:render}. This phase transition can occur rapidly and locally due to temperature changes which may be induced by a laser, for optical storage, or electrically, for non-volatile random access memory~\cite{ren2011phase, rios2015integrated, lai2003current, le2020overview}.  Currently, the most commonly used PCM for memory applications is Ge$_{2}$Sb$_{2}$Te$_{5}$ (GST)~\cite{ren2011phase, le2020overview}, and hence there is wide interest to better characterize this material both experimentally and through atomistic modeling. Our focus is on developing an accurate and efficient computational model for atomistic simulations of GST for use by the broader community.

There are a range of approaches to model the interactions of the constituent atoms of a material via computer simulation. Empirical potentials that can be described by simple analytic functions of the particle coordinates, such as the ubiquitous Lennard-Jones or Coulombic potentials, can often successfully capture material properties.  The simplicity of such models facilitate rapid computation of forces and energies, allowing for material simulations at large length and time scales. While these simple models are successful in many cases, they are insufficient to model a variety of materials, particularly when the electronic structure plays an important role in the interatomic interactions. This challenge arises for many phase change materials, such as GST, which are known to exhibit competing structural motifs~\cite{caravati2007coexistence}. Consequently, methods to incorporate electronic structure, such as Density Functional Theory (DFT), are often used to generate the forces and energies needed for a dynamical simulation. This procedure is referred to as \textit{ab initio} molecular dynamics (MD), and has been shown to reproduce a number of the experimentally known properties for GST~\cite{kalikka2016crystallizatiom}. As pointed out in recent work~\cite{mocanu2018modeling, zhou2023device}, the computational expense of \textit{ab initio} MD effectively precludes exploration of the time and length scales needed to study macroscopically relevant properties, such as the phase change cycle of GST. Thus, there is a need to explore alternative methods that combine accuracy and computational efficiency.

Machine learning (ML) offers an approach that is being rapidly adopted to train computationally efficient models that reproduce the potential energy surface of DFT-based calculations~\cite{friederich2021machine, ori2019chalcogenide}.  The field of ML potentials has been growing since 1992, beginning with general neural network (NN) potentials that draw on existing ML methods to train models on large sets of DFT data~\cite{friederich2021machine}. Since then, regression models, such as the Gaussian Approximation Potential (GAP) have become increasingly popular, where the training process fits preexisting, high-dimensional functions that enforce physical symmetries such as translation, rotation, and permutation of atoms of the same type~\cite{kolb2016permutation, li2019critical, friederich2021machine, behler2007generalized, behler2016perspective}. Recently, several ML potentials that are directly trained on GST configurations computed with DFT have been developed: El Kheir {\it et al.}~\cite{kheir2024unraveling} and Choi {\it et al.}~\cite{choi2024study} developed efficient NN-based models, and  Zhou {\it et al.}~\cite{zhou2023device} developed a GAP~\cite{GAP, mocanu2018modeling}. 
%These models successfully reproduce DFT forces and energies at a fraction of the computational cost. 
Despite the tremendous improvement in computational performance compared to DFT, the complexity of the GAP framework still makes it a challenge to simulate the length and time scales necessary to study the phase change properties that make GST so valuable.

Here  we consider an alternate ML approach that is potentially even more computationally expedient. 
The Atomic Cluster Expansion (ACE)~\cite{bochkarev2022efficient, drautz2019atomic, lysogorskiy2023active}, a recently introduced regression model, is an efficient alternative with the flexibility to represent many existing empirical and ML potentials~\cite{lysogorskiy2021performant}. Thus, our focus is to use the ACE framework to train an accurate and efficient GST potential.

In order to train a suitable ML potential, it is critical that the training data set contains sufficient and accurate information on the interactions and states that are of interest to model~\cite{friederich2021machine}. Since ML potentials are often trained for systems with complex quantum interactions, DFT offers clear advantages as a source of training data. This process of training a model on DFT interactions is known as direct learning, and has made up the vast majority of ML potentials to date. However, in order to build a representative training set, thousands of DFT calculations may be needed to cover a wide range of structures and thermodynamic conditions. Thus, the high computational cost of DFT presents a current limitation of this direct learning approach~\cite{ori2019chalcogenide}.

As an alternative to this traditional approach, a method of indirect learning has recently been proposed~\cite{morrow2022indirect}. Indirect learning is the practice of training a potential using an intermediate ML potential which was itself trained directly on the DFT; hence the moniker ``indirect learning''. By generating the training set using an ML potential, expensive DFT calculations can be bypassed, allowing for a much larger dataset to be generated than would be plausible using direct learning. Morrow et al.~\cite{morrow2022indirect} suggest that this procedure would allow the training set to sample a wider range of conditions than would be possible if it was generated using DFT, which may lead to a ML potential that is transferable to a broader range of conditions.  

In this paper, we develop potentials for GST using the ACE framework where we consider both direct learning using the published DFT training set of the recently developed GAP model for GST~\cite{zhou2023device} as well as indirect learning from additional training configurations generated by this GAP model. (Note that the DFT training set of ref.~\citenum{kheir2024unraveling} is also freely available, though it had not been published at the time this work was done).
The resulting ACE potentials  reproduce the behavior of GAP at approximately 1000 times the computational speed on equivalent computational resources. We  validate that the properties of the indirectly learned potential agree with both GAP and the directly learned potential across a wide range of conditions. We then expand the training set used to do indirect learning and discuss the implications of our resulting potential. Finally, we demonstrate the ability of our potential to efficiently model the phase change behavior of GST in device-scale simulations of $10^6$ atoms.

\section{Methods}

\begin{table*}
\caption{Brief summary of the potentials presented in this paper.}
\begin{tabular}{ |>{\centering\arraybackslash}p{1.5cm}||>{\centering\arraybackslash}p{2.5cm}|>{\centering\arraybackslash}p{2.8cm}|>{\centering\arraybackslash}p{9.9cm}| }
 \hline
 %\multicolumn{4}{|c|}{Potentials Discussed} \\
 %\hline
 Potential & Training Style & Atom Environments & Training Set Description\\
 \hline
 \dirGAPTS & Direct from DFT & 340,709 & GAP Training Set \\
 \indirGAPTS & Indirect from GAP & 340,709 & GAP Training Set \\
 \indirEXP & Indirect from GAP & 1,441,445 & GAP Training Set and Environments from 12 Isochoric Temperature Sweeps \\
 \hline
\end{tabular}
\label{table:potentials}
\end{table*}

\subsection{\label{sec:framework}Potential Framework and Specifications}

We train all ML potentials using the Atomic Cluster Expansion (ACE)~\cite{bochkarev2022efficient, drautz2019atomic, lysogorskiy2023active}  framework which provides a complete set of many-body basis functions to describe interatomic interactions due to an atom's local environment. ACE models are fitted to interatomic force and energy data using the Performant implementation of the Atomic Cluster Expansion (PACE)~\cite{lysogorskiy2021performant}.

In the ACE framework, the local environment of an atom $i$ is described by a collection of atomic properties $\varphi_i^{(p)}$, where $p$ is the property index. $\varphi_i^{(p)}$ is a sum of $\nu$-body terms up to a maximum of $\mathbf{\nu}_{max}$ bodies, where we select $\mathbf{\nu}_{max}=6$. For each such term $\nu$, the interactions are described by a finite sum of $\nu$-body basis set functions $\mathbf{A}_{i \mu \mathbf{nlm}}$ with weights $\Tilde{c}_{\mu_i \mu \mathbf{nlm}}^{(p)}$. Formally,  atomic property $p$ of atom  $i$ is given by 
\begin{equation}
    \mathbf{\varphi}_{i}^{(p)} = \sum_{\nu=2}^{\nu_{max}} \sum_{\mathbf{\mu} \mathbf{nlm}} \Tilde{c}_{\mu_i \mathbf{\mu} \mathbf{nlm}}^{(p)}  \mathbf{A}_{i \mu \mathbf{nlm}}
\label{eq:atomic-property}
\end{equation}
where $\mathbf{\mu} = (\mu_1, \mu_2, ... , \mu_{\nu})$ corresponds to the chemical species, while $\mathbf{n}$, $\mathbf{l}$, and $\mathbf{m}$ are vectors of length $\nu-1$ representing basis function indices; the sum over $\nu$ starts with 2-body terms. The ACE potentials are trained using 1200 independent basis functions for each element Ge, Sb, and Te. Here, these many-body basis functions $\mathbf{A}_{i \mu \mathbf{nlm}}$ are products of $\nu$ atomic bases $A_{i \mu_{t} n_{t} l_{t} m_{t}}$, which are each built by summing a single-particle basis function $\phi_{\mu_i \mu_j n l m}$ over the $j$ nearest neighbors to atom $i$,
\begin{equation}
    \mathbf{A}_{i \mathbf{\mu} \mathbf{nlm}} = \prod_{t=1}^{\nu-1} A_{i \mu_{t} n_{t} l_{t} m_{t}} = \prod_{t=1}^{\nu-1} \left( \sum_{j} \delta_{\mu \mu_j} \phi_{\mu_i \mu_j n l m}(\mathbf{r}_{ji}) \right).
\end{equation}
The single-particle basis function is given by $\phi_{\mu_i \mu_j n l m}=R_{n l}^{\mu_i \mu_j}(r_{ji})Y_{lm}({\displaystyle {\hat {\mathbf {r} }}}_{ji})$, where $Y_{lm}({\displaystyle {\hat {\mathbf {r} }}}_{ji})$ is a spherical harmonic, and $R_{n l}^{\mu_i \mu_j}(r_{ji})$ is a radial function which is made up of a complete radial basis; specifically, we use a radial basis of exponentially-scaled Chebyshev polynomials given by
\begin{equation}
    g_k(r)=\frac{1}{4}(1-T_k(x))\left(1+\cos\left(\frac{\pi r}{r_{cut}}\right) \right)
\end{equation}
where $T_k(x)$ are the Chebyshev polynomials, expressed in terms of $x=1-2(\frac{e^{-\lambda (r/r_{cut}-1)}}{e^{\lambda}-1})$. For all potentials considered, the outer cutoff radius $r_{cut}=5.5$~\AA\ is chosen to match the cutoff of the GAP potential, and the parameter $\lambda=5.25$ following ref.~\citenum{bochkarev2022efficient}. Additionally, we implement an inner cutoff of 1~\AA\  of the ML potential where an empirical radial core repulsion potential takes over, 
\begin{equation}
    U_{\text{rep}}= U_0  \frac{e^{- \sigma r^{2}}}{r}
\end{equation}
with $U_0= 100\, \text{eV} \cdot \text{\AA}$ and $\sigma= 5$~\AA$^{-2}$, the default values of PACE~\cite{lysogorskiy2021performant}.

Because the basis functions $\mathbf{A}_{i \mu \mathbf{nlm}}$ do not have rotational and inversion symmetries, the training process is instead done using a separate expansion of basis functions $\mathbf{B}_{i \mu \mathbf{nlm}}$ that do possess these symmetries.  The training results in weights $c_{\mu_i \mu \mathbf{nlm}}^{(p)}$ in the $\mathbf{B}_{i \mu \mathbf{nlm}}$ basis, which then are then transformed into the $\mathbf{A}_{i \mu \mathbf{nlm}}$ basis with corresponding weights $\Tilde{c}_{\mu_i \mu \mathbf{nlm}}^{(p)}$.  The evaluation of forces and energies is more efficient in the basis of $\mathbf{A}_{i \mu \mathbf{nlm}}$. The transformation from  $c_{\mu_i \mu \mathbf{nlm}}^{(p)}$ to $\Tilde{c}_{\mu_i \mu \mathbf{nlm}}^{(p)}$ is done by multiplication with the generalized Clebsch-Gordan coefficients, ensuring that the expansion given by eq.~\ref{eq:atomic-property} is invariant under rotation and inversion.

Training of the ACE potential is done by minimizing a loss function which is necessarily designed to penalize discrepancies between both the forces and energies predicted by the model, and those provided in the reference dataset. A parameter $\kappa$ allows the relative weights placed on energies and forces in the training process to be adjusted: the force component of the loss function has a weight of $\kappa$, while the energy component has a weight of $1-\kappa$. All ACE potentials discussed in this work are trained with $\kappa=0.3$, weighting energies higher than forces;  the errors for forces and energies (discussed in sec.~\ref{sec:errors}) are nearly invariant for $0.3\le \kappa \le 0.7$. In all training runs, we randomly select 80\% of configurations to train the potential and the remaining 20\% make up the test set to validate the performance on new environments.

An ACE model may be defined in terms of one or many atomic properties $(\mathbf{\varphi}_i^{(1)},\mathbf{\varphi}_i^{(2)},...)$, on which energy may be linearly or non-linearly dependent. In a simple linear model, the energy of atomic environment $i$ is linearly dependent on a single atomic property $E_{i}=\mathbf{\varphi}_i^{(1)}$. 
While this linear model is the most computationally efficient, it has been shown that convergent results can be reached using fewer body-order terms when an additional non-linear contribution is introduced~\cite{bochkarev2022efficient}. Since the complex behaviors of phase-change materials are dependent on many higher body-order corrections, we include a non-linear energy dependence on atomic properties, namely
\begin{equation}
    E_{i}=\mathbf{\varphi}_i,^{(1)}-\sqrt{\mathbf{\varphi}_i^{(2)}}.
\end{equation}
In this model, energy depends linearly on one atomic property and nonlinearly on a second, taking the form of the Finnis-Sinclair potential for transition metals~\cite{FS1984}.

Note that, since our aim is implementation rather than development of ACE,  we have omitted details that are not needed to reproduce our findings, including a thorough description of the generalized Clebsch-Gordan coefficients and the formal definition of the loss function.  Please refer to  Bochkarev et al.~\cite{bochkarev2022efficient} and Lysogorskiy et al.~\cite{lysogorskiy2021performant} for a more detailed discussion. 

\subsection{\label{sec:dataset}Dataset}

\begin{figure*}
\includegraphics[width=1.0\textwidth,keepaspectratio]{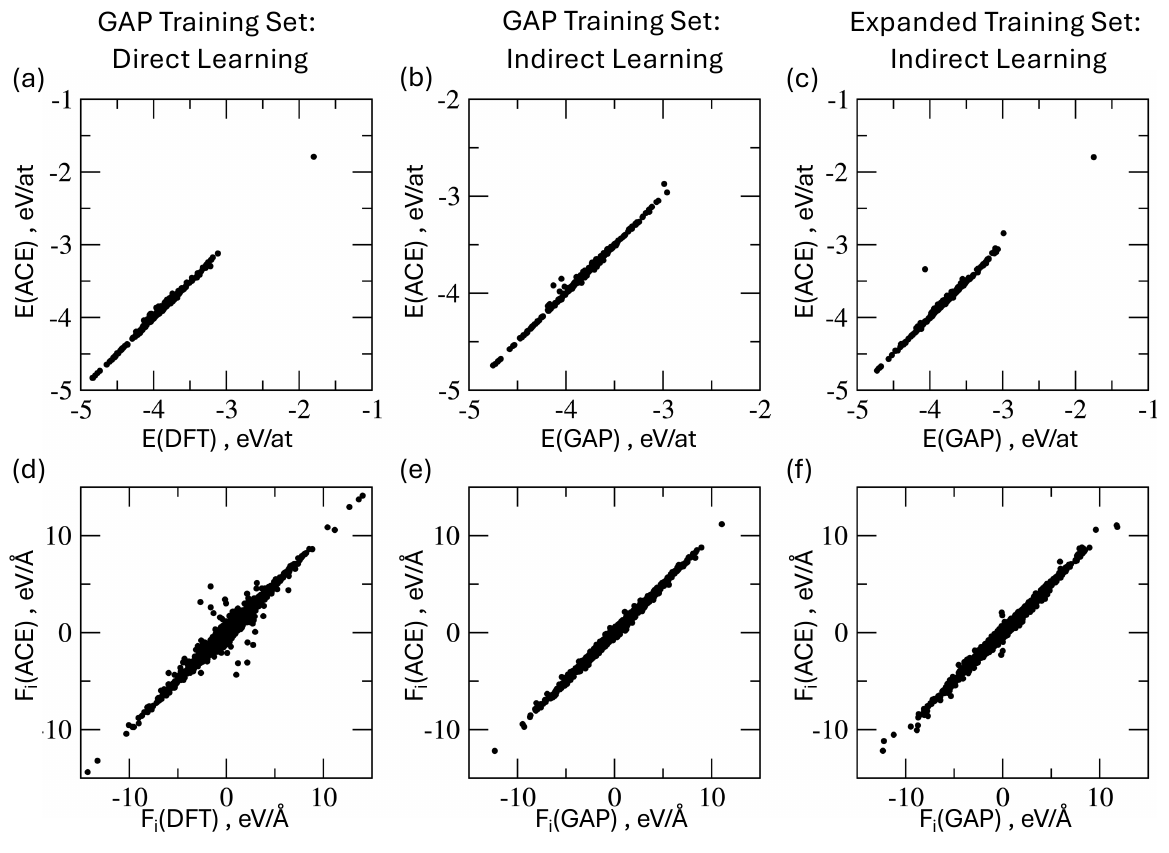}
\caption{Energy and force parity plots for the three trained potentials on their random 20\%\ test set: energy predictions compared between (a) DFT and \dirGAPTS, (b) GAP and \indirGAPTS, and (c) GAP and \indirEXP; force component predictions compared between (d) DFT and \dirGAPTS, (e) GAP and \indirGAPTS, (f) GAP and \indirEXP.}
\label{fig:parity_plots}
\end{figure*}

When Zhou {\it et al.}~\cite{zhou2023device} published their GAP GST potential, they also published its stoichiometrically and structurally diverse training dataset, with corresponding forces and energies computed by DFT. This dataset consists of elemental, binary, and ternary (Ge$_2$Sb$_2$Te$_5$) compositions. The dataset includes dimers, amorphous and partially amorphous structures, and a variety of crystal structures. Rather than creating a new training set, we use this thorough dataset to train ACE GST potentials. Moreover, this provides consistency in the training set between the GAP model and our ACE potential to facilitate comparisons.  Note that we remove dimer configurations and instead implement an arbitrary core repulsion in the training process; we exclude dimers because attempts to include  both dimer and bulk GST configurations in PACE training biases pairwise interactions towards higher energies, leading to unstable performance at higher temperatures in condensed phases.  Furthermore, dimers present a very different local environment than bulk condensed phases which we are interested in simulating. As a result, we exclude 210 dimer configurations and ultimately train on a subset of the GAP training set consisting of 340,709 atom environments. 

As a control case, an ACE potential is first trained directly on this DFT data for forces and energies. For indirect training, we use the same environments but use the LAMMPS-MD simulator~\cite{LAMMPS, QUIP} and the Atomic Simulation Environment~\cite{ASE} (ASE) to recompute the forces and energies with the GAP potential, and train an ML potential on this recomputed data set. For brevity, we will refer to the directly learned version of this potential as \dirGAPTS\ and the indirectly learned version as \indirGAPTS. A thorough comparison of these potentials, trained on the same set of configurations, is included in the results section.

We take advantage of the low computational cost of GAP relative to DFT to expand our training set using indirect learning.  Specifically, we use the LAMMPS implementation of GAP to generate 504-atom configurations of GST at densities in the experimentally relevant range between 5.85 g/cm$^3$ and 6.40 g/cm$^3$, in increments of 0.05 g/cm$^3$. For each isochore, systems are cooled from 1200~K to 300~K over 300~ps (with a timestep of 1~fs), and configurations are saved at intervals of 5~K for later training. Note that, at this cooling rate, all simulations remain in liquid or amorphous states. We add these configurations to the existing GAP dataset, resulting in a training set of 1,441,445 atom environments that incorporates both diverse structures/stoichiometries, and many equilibrated structures at thermodynamic conditions of interest for simulation. An ACE potential is trained on the expanded training set, which we refer to as \indirEXP.  Table~\ref{table:potentials} provides a brief summary of each trained potential we discuss.

\subsection{\label{sec:errors}Errors}

For the potentials trained directly and indirectly on the GAP training set, we reach test set root-mean-square errors (RMSE) for the energy of 19.92 meV/atom and 19.23 meV/atom respectively, and RMSE for the force components of 148.95 meV/\AA\ and 88.01 meV/\AA. The potential trained on the expanded (indirect) training set achieved an energy RMSE of 26.90 meV/atom and a force RMSE of 91.48 meV/\AA. These error metrics are within the range reported in prior works~\cite{zhou2023device, kovacs2021linear} where they find energy RMSEs on the order of 1-10 meV/atom and force RMSEs on the order of 10-100 meV/\AA. Thus our trained ACE potentials can be expected to predict the forces and energies of configurations not included in training with similar reliability. Parity plots of forces and energies are shown in Fig.~\ref{fig:parity_plots}. Though we do note some deviations from perfect agreement, the results of this paper demonstrate outstanding agreement of the thermodynamic and structural properties of these potentials with GAP and experiments under a range of conditions. We consider these practical measures to be a more useful indicator of the utility of the model than specific error metrics for energies and forces that are not experimentally accessible.  It is also worth noting that the directly learned potential exhibits a significantly higher force RMSE than either of the indirectly trained potentials, though this is expected since ML potentials enforce smoothing of the potential energy surface, which is already incorporated in the GAP training configurations but not in the DFT configurations~\cite{mueller2020machine}.

\vspace{\baselineskip}

\section{Results}

\subsection{\label{sec:dir-vs-indir}Comparing Direct and Indirect Learning}
\begin{figure}
\centering
\includegraphics[width=0.5\textwidth,keepaspectratio]{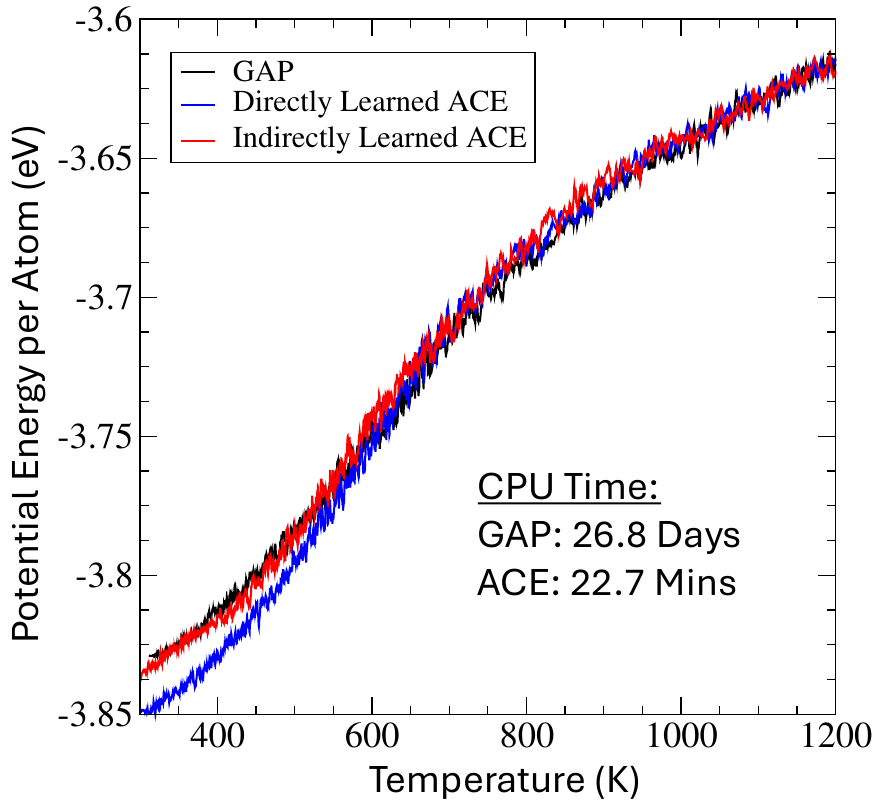}
\caption{Potential energy versus temperature during a rapid quench at constant density for GAP, \dirGAPTS, and \indirGAPTS. The differences in potential energy at low temperature are a consequence of the  system falling out of equilibrium at this cooling rate. }
\label{fig:PE-sq}
\end{figure}

\justifying

Our results begin with an overview of the effectiveness of the training process of the ACE potentials, both for the case of direct learning from DFT and indirect learning from GAP. We then explore the effects of expanding the training set, and conclude by demonstrating the ability of our ACE potential to efficiently reproduce a variety of GST behaviors either predicted by the GAP potential or observed experimentally.

We first compare the behavior of \dirGAPTS, trained directly on DFT, and \indirGAPTS, trained indirectly using the same configurations but with forces and energies evaluated by the GAP potential. To begin, we compare the potential energy of these models along an isochoric (fixed density) path. We focus here on amorphous states by performing  rapid quenches from 1200~K to 300~K over 300 ps at a density of 6.00~g/cm$^3$; crystallization occurs with slower cooling rates which will be addressed later. Figure~\ref{fig:PE-sq} shows that the potential energies of \dirGAPTS\ and \indirGAPTS\ agree quantitatively across the temperature range of interest for simulation. We note that, at this rapid cooling rate, systems fall out of equilibrium around 500~K, leading to cooling rate-dependent discrepancies in the potential energy at the lowest temperatures. These results indicate that, for the training set considered, indirect learning does not sacrifice accuracy in predicting potential energy compared with direct learning. In addition to showing agreement between \dirGAPTS\ and \indirGAPTS, we also include potential energy data for an identical thermodynamic path simulated using the GAP potential. This demonstrates the ability of \dirGAPTS\ and \indirGAPTS\ to match the results of the existing GAP potential on new environments that were not included in the training set. 

We also emphasize the computational efficiency of the ACE model. Using 8 nodes of a high-performance computing cluster to simulate 504 atoms of GST, this GAP run took nearly 27 days to run, while the ACE runs needed less than 23 minutes using the same resources: a speedup by a factor of about 1700. The precise value of the speedup varies based on the specific hardware, but we consistently observe a speedup by three orders of magnitude, which opens the possibility of considering a much broader set of physical questions.  A more detailed review of the computational performance of our ACE potential, including GPU acceleration, is discussed later in sec.~\ref{sec:Speed}.

\vspace{\baselineskip}

\begin{figure*}
\centering
\includegraphics[width=1.0\textwidth,keepaspectratio]{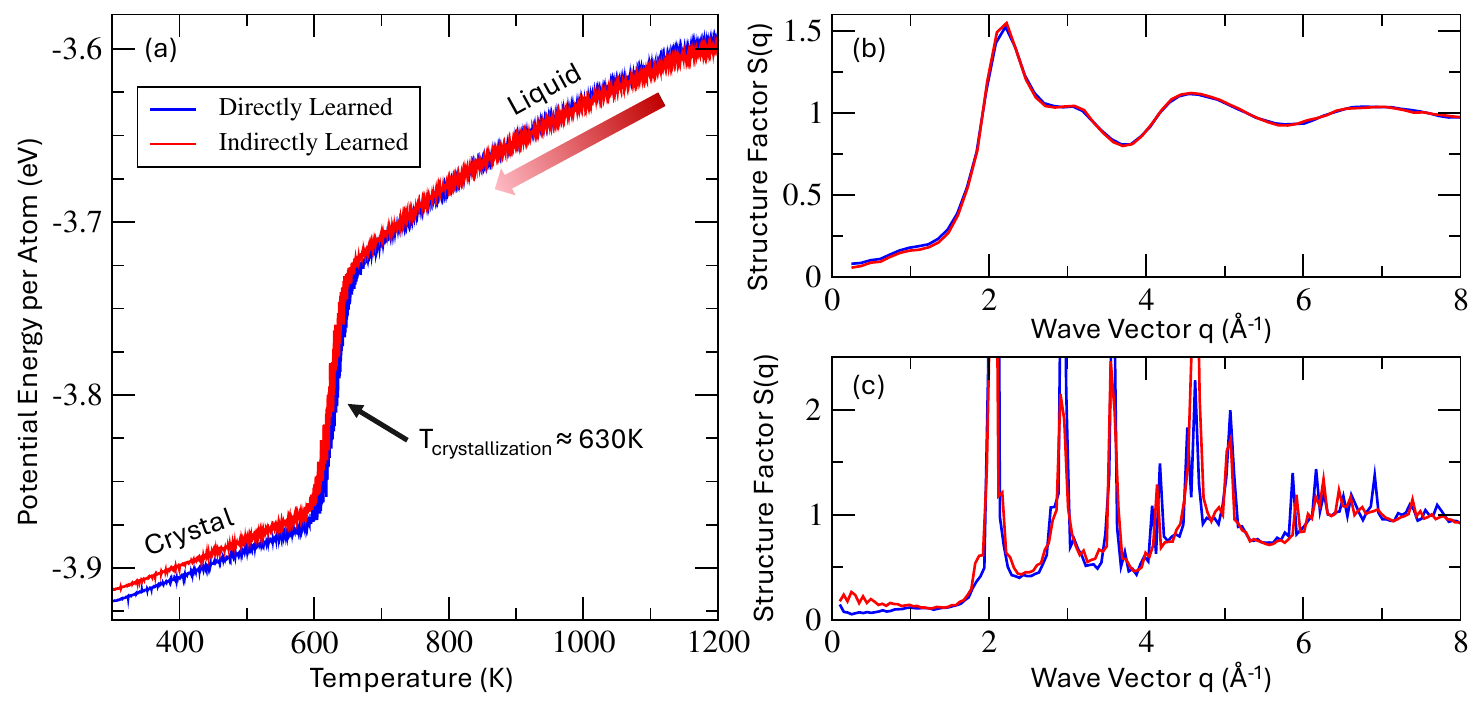}
\caption{Comparisons between the directly learned and indirectly learned potentials, \dirGAPTS, and \indirGAPTS, trained on the same training set. (a) Potential Energy as a function of temperature during a 40 ns isobaric quench at 1 bar. The potential energy drops sharply around 630~K due to crystallization. (b) A comparison of the structure factor in the liquid state at 900~K and (c) the crystal state at 600~K at (both at P=1 bar). Both show strong agreement.}
\label{fig:Isobar}
\end{figure*}

\justifying

To further probe the capabilities and the level of agreement between the directly and indirectly learned potentials, we compare the behavior of \dirGAPTS\ and \indirGAPTS\ along isobaric (rather than isochoric) temperature sweeps. We cool at a slower rate to minimize any effects of falling out of equilibrium, which also opens the possibility of spontaneous crystallization. We simulate GST systems of 1080 atoms cooled at a constant $P=1$~bar from $T=1200$~K to 300~K over 40~ns. It is worth noting that a simulation of this duration (250 times as many particle updates as the rapid quench we previously discussed) is readily accessible due to the speed of the ACE model, and that simulating such a system size and duration would be challenging using the GAP potential without employing much larger computational resources, and likely impossible using DFT. Figure~\ref{fig:Isobar}(a) shows the $T$-dependence of the potential energy; in both cases, there is a liquid-to-crystal phase transition indicated by the sharp drop of potential energy at $T \approx 630$~K. Moreover, the values of the potential energies of \dirGAPTS\ and \indirGAPTS\ are in quantitative agreement throughout the liquid and crystal phases, showing that even nontrivial information about the phase transition is preserved when switching from direct to indirect learning (for this choice of training data). We note that there is a slight difference in the potential energy in the crystal state, which results in part from random differences that occur in the crystal nucleation process; these stochastic effects can also lead to variation in the precise temperature at which crystallization occurs.

As a more sensitive test of how well the \dirGAPTS\ and \indirGAPTS\ potentials match each other, we compare the liquid and crystal GST structures predicted by these models. We prepared GST systems at 900~K and fixed $P=1$~bar in the liquid state, and at 600 K and $P=1$~bar where the system spontaneously crystallizes. Figure~\ref{fig:Isobar}(b) shows that the structure factor 
\begin{equation}
    S(q)= \frac{1}{N} \langle \rho(\mathbf{q})\rho^\star(\mathbf{q}) \rangle
\end{equation}
in the liquid state of systems simulated using \dirGAPTS\ and \indirGAPTS\ is nearly indistinguishable; here, $\rho(\mathbf{q})$ is the Fourier transform of the density at wave vector $\mathbf{q}$, and $S(q)$ is averaged over vectors with the same magnitude $q=\vert \mathbf{q} \vert $. Figure~\ref{fig:Isobar}(c) shows the same information for the spontaneously nucleated crystal phase. Since we choose vectors with magnitudes discretized by $\delta q = 2\pi/L$ to guarantee periodicity, we crystallize a somewhat larger system of 8064 atoms so that the  resolution $\delta q$ is fine enough to capture all peak locations. The peak locations of the spontaneously formed crystals obtained using \dirGAPTS\ and \indirGAPTS\ match each other, suggesting agreement between the crystal lattice structures predicted by these potentials. We later compare both liquid and crystal structure factors with experiments.

In short, using the previously published extensive training set, we find quantitative agreement between ACE models trained either directly from DFT forces and energies, or indirectly from GAP forces and energies. Apparently, indirect learning can be just as effective as direct learning, provided the indirect model has been well parameterized and an adequate data set is used for training. We next consider expanding this training set.

\subsection{\label{sec:expanded-set}Expanded Training Set}

\begin{figure}[b]
\centering
\includegraphics[width=.5\textwidth,keepaspectratio]{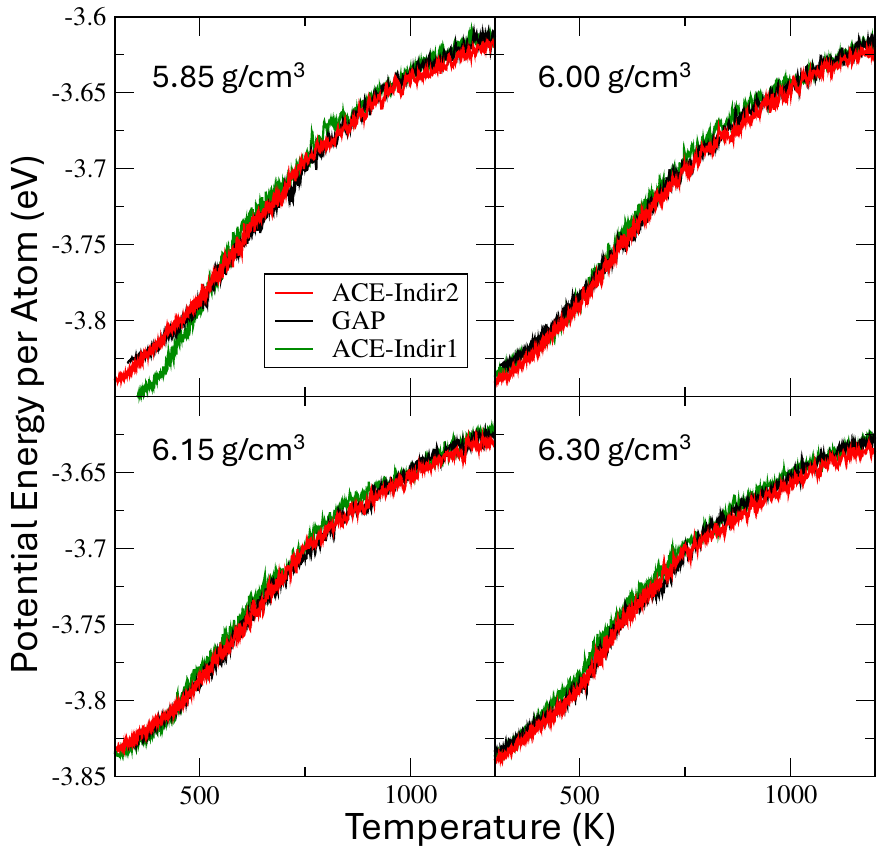}
\caption{Potential Energy as a function of temperature during 300 ps isochoric quenches at varied densities compared between GAP and both indirectly learned ACE potentials.
%The data for all potentials match, except at the lowest temperatures where cooling-rate dependent effects become significant.
}
\label{fig:Isochore}
\end{figure}

\justifying

\begin{figure*}
\centering
\includegraphics[width=1.0\textwidth,keepaspectratio]{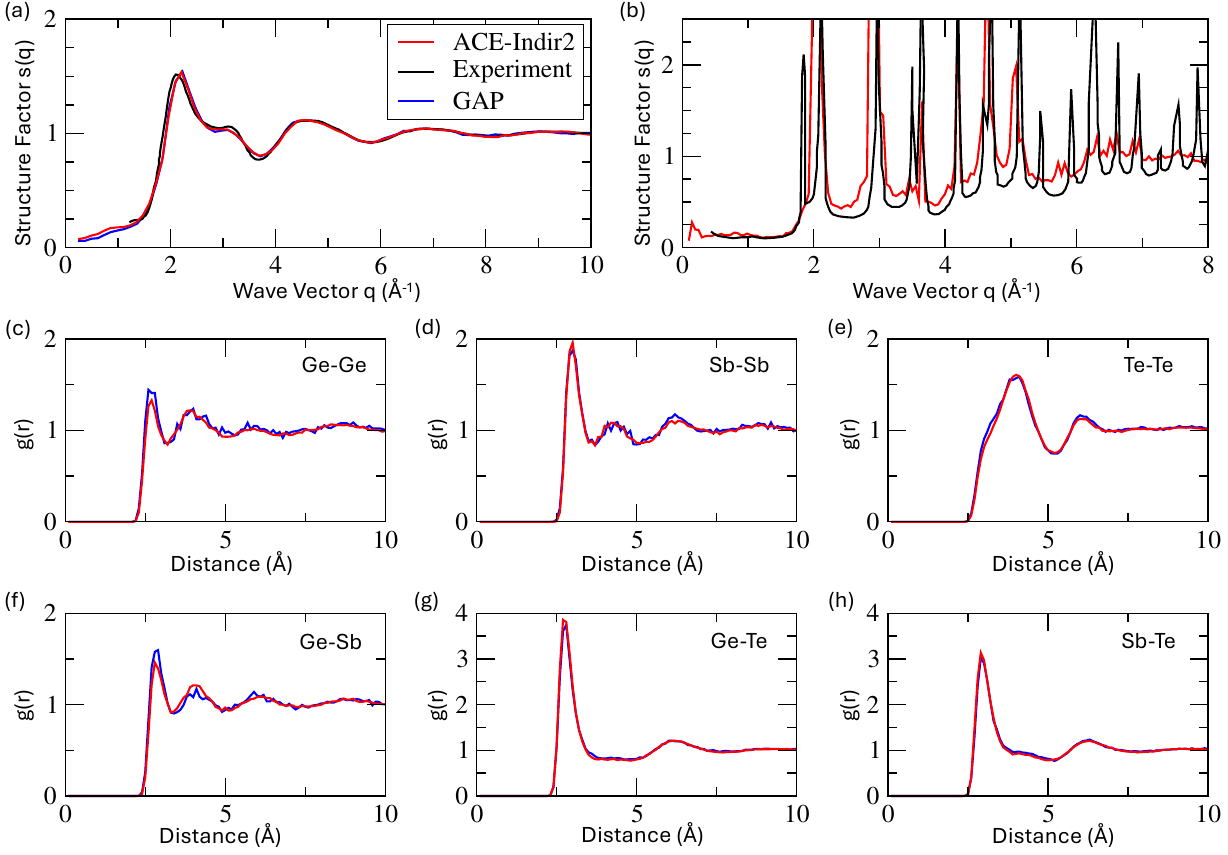}
\caption{A structural analysis of equilibrated GST structures, including comparison with the GAP potential and experiment~\cite{kohara2006sq}: (a) The liquid structure factor $S(q)$ of the ACE potential compared with GAP and experiment at $T=900$~K and (b) the crystal $S(q)$ compared between ACE and experiment at $T=550$~K. (c-h) The partial radial distribution functions of each element pair, compared between ACE and GAP.  In all panels, red is \indirEXP, blue is GAP, and black is experiment.}
\label{fig:Validation}
\end{figure*}

\justifying

Having established the correspondence between the directly and indirectly learned potentials trained on a common set of configurations, we now consider if the ML potential can be further refined or improved by expanding  our training set.  Indirect learning makes this expansion feasible due to the improved speed of GAP compared to DFT.  Specifically, we will consider 1,441,445 atom environments (compared to the original 340,709 environments) by incorporating configurations from 12 isochoric quenches of 504 atoms along with the existing training set used to train the GAP model. These isochoric data expand the thermodynamic states sampled in the training process, and thus might result in an ACE potential that is more transferable across the phase diagram.

We test the agreement between the ACE and GAP models by using this newly trained ACE potential -- \indirEXP\ -- to reproduce the 12 isochoric simulations included in the expanded training set.  Specifically, Fig.~\ref{fig:Isochore} shows the potential energy plotted against temperature for 4 representative isochoric quenches at densities 5.85 g/cm$^3$, 6.00 g/cm$^3$, 6.15 g/cm$^3$, and 6.30 g/cm$^3$. We  see that the indirectly learned ACE potential trained on the expanded training set matches -- qualitatively and quantitatively -- the energy calculations of the GAP potential on which it was trained. 

Also included in Fig.~\ref{fig:Isochore} is the potential energy calculated by \indirGAPTS\, trained on only the GAP training set. The \indirGAPTS\ potential matches both the expanded ACE potential and the GAP training data. Though we anticipated that explicitly including additional isochoric quenches in the training process might improve performance in these regions of the phase diagram, it is evident that the GAP training set was already large and diverse enough to reach consistent results using the ACE model. This suggests that the quality of the training set may be more important than choosing direct versus indirect learning. That said, it is easy to imagine a situation in which a potential is trained without such an extensive pre-existing dataset.  The fact that we have shown the ability to reach similarly performing potentials using direct and indirect learning could drastically reduce the number of expensive DFT calculations needed. We can also illustrate the importance of a diverse training set by training an ACE potential using \textit{only} the configurations from the 12 isochoric training samples. Such a trained potential is completely unparameterized on crystal environments, and as might be expected, the potential does not match the behavior of GAP and experiment over as broad a range of thermodynamic conditions as potentials that are trained using the original GAP dataset.

\begin{figure}[b]
\centering
\includegraphics[width=0.5\textwidth,keepaspectratio]{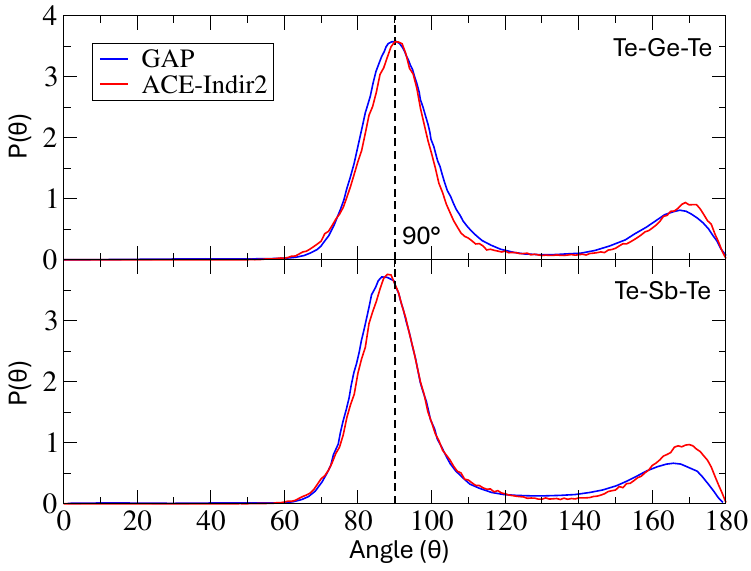}
\caption{The bond angle distribution $P(\theta)$ for Te-Ge-Te bonds and Te-Sb-Te bonds, compared between \indirEXP\ calculated here and GAP results from ref.~\citenum{mocanu2020quench}.}
\label{fig:BAD}
\end{figure}

\begin{figure*}
\centering
\includegraphics[width=1.0\textwidth,keepaspectratio]{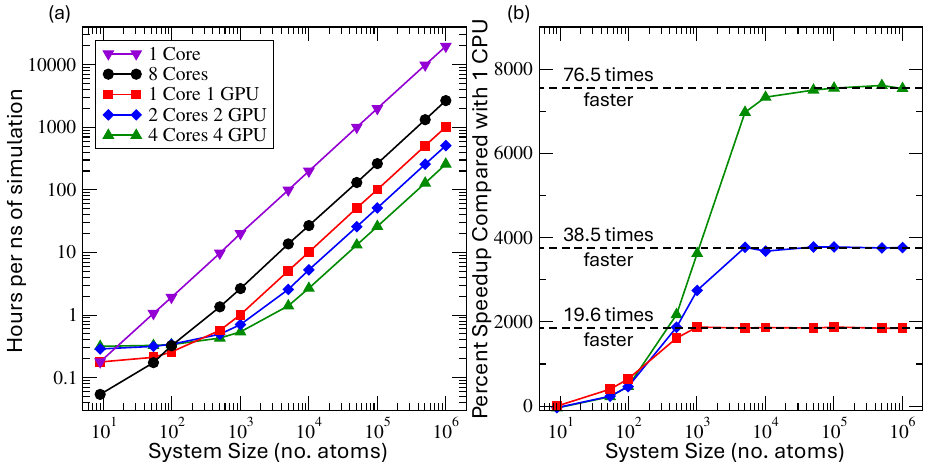}
\caption{The speed of the ACE model as tested on a variety of computing configurations, all using local resources. %without significant barriers to access. 
(a) The performance of the potential (hours to complete 1~ns of simulation at various system sizes) for calculations using 1 and 8 CPU cores, as well as 1, 2, and 4 GPUs. (b) Using the case of 1 CPU core as a reference, the GPU acceleration available using 1, 2 and 4 GPUs. Note that GPU acceleration is optimized at system sizes on the order of $10^3$-$10^4$ atoms or larger. The speedup compared to 8 CPUs (the configuration used for most simulations in this manuscript) is obtained roughly by dividing these results by 7.5.}
\label{fig:GPU}
\end{figure*}

\justifying

\subsection{\label{sec:validation}Potential Validation}

We have already provided evidence for the correspondence between our indirectly learned ACE potentials and the GAP potential on which they were trained. This section presents a more thorough analysis of our ACE model's agreement with GAP and, more importantly, experimental data. Though we showed that \indirGAPTS\ and \indirEXP\ agree energetically, we focus on the \indirEXP\ potential for the remainder of this section.

For this potential to serve as a representative model of the amorphous and crystalline phases of GST, it should reproduce structures that agree with the GAP potential and experimental results; we demonstrate here that this is the case. Figure~\ref{fig:Validation}(a) shows the structure factor $S(q)$ of liquid GST simulated at 900~K and $P=1$~bar using the GAP and \indirEXP\ potentials, as well data from an experiment~\cite{kohara2006sq}. These $S(q)$ data are nearly indistinguishable from each other.  We can make a more refined comparison of the structure by evaluating the partial radial distribution functions (RDF) $g(r)$, for each pair of elements Ge, Sb, Te, as shown in  Figure~\ref{fig:Validation}(c-h). From both the structure factor and partial RDFs, it is clear that our ACE model reproduces the amorphous structure predicted by GAP, with even individual element pairs showing the predicted distribution. Additionally, the computed structure very nearly matches the experimentally measured structure, indicating the ability for \indirEXP\ to provide insight into experimentally observed behaviors. 

In addition to the liquid phase, we examine the structure factor of the crystal that spontaneously nucleates using our ACE potential. As expected from experiments, the spontaneously formed crystal at $T=550$~K and $P=1$~bar has a cubic structure~\cite{nonaka2000crystal}.  Figure~\ref{fig:Validation}(b) compares the structure factor from \indirEXP\  with the experimental crystal structure factor~\cite{kohara2006sq}. Note that data for the spontaneous crystallization using the GAP potential is omitted due to the computational expense required for such a simulation. We again crystallize a 8064 atom system of GST  to ensure a sufficient resolution of the structure factor; the $S(q)$ peak locations of the crystal formed using \indirEXP\ align with those of the experimental structure factor.

To complete our comparison of the models prediction of GST's structure for common metrics, we calculate the coordination numbers and bond angle distributions $P(\theta)$ for the amorphous state at $T=300$~K. In amorphous GST, the \indirEXP\ predicts an average coordination number around Germanium (with a cutoff of 3.15~\AA) of 4.4, with corresponding values for Sb and Te of 3.7 and 3.0. This compares well with DFT studies, which report amorphous GST coordination numbers around Ge, Sb, and Te of 4.2, 3.7, and 2.9, respectively~\cite{akola2012amorphous}. The distribution of bond angles ($P(\theta)$) around central atoms provides a more nuanced picture of local order.  Figure~\ref{fig:BAD} shows $P(\theta)$ in the same amorphous GST for the cases of Te atoms around a central Ge or Sb atom along with published GAP simulation data~\cite{mocanu2020quench}. $P(\theta)$ predicted by this potential corresponds to the expected behavior, with both cases showing a large peak at $\theta=90^\circ$ and a smaller bump around $\theta=170^\circ$, consistent with the locally cubic ordering of amorphous GST.

\begin{figure*}
\centering
\includegraphics[width=1.0\textwidth,keepaspectratio]{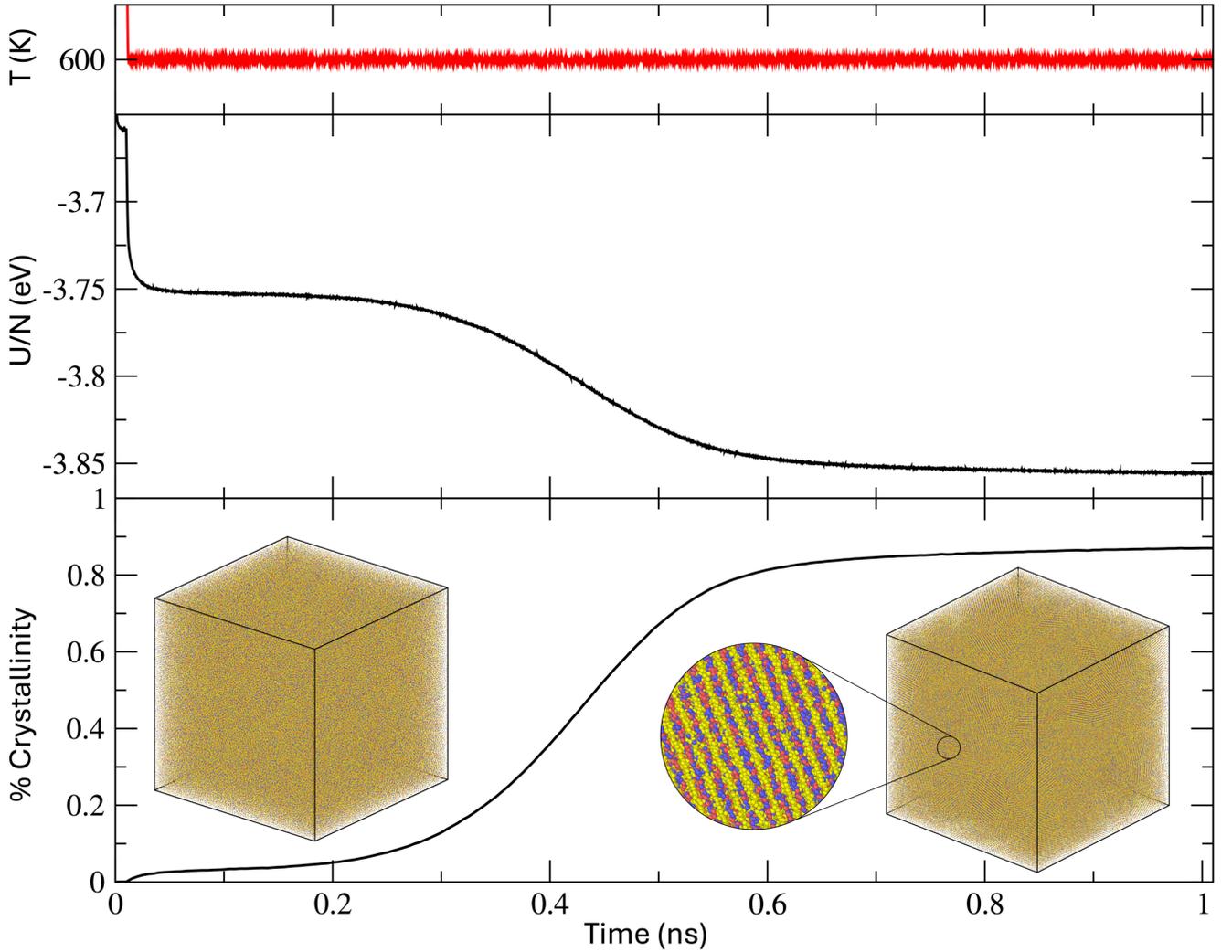}
\caption{Evolution of a 1,008,000 atom GST system as the liquid crystallizes at constant $T=600$~K. The potential energy per atom and percent of the system identified as crystalline are plotted over the 1~ns phase transition, where several crystal nuclei form and grow throughout the system. The graphic shows the 1,008,000-atom system in each of the two phases; Ge are shown in red, Sb are shown in blue, and Te are shown in yellow.}
\label{fig:million}
\end{figure*}

\subsection{\label{sec:Speed}Speed}

Having shown extensive structural correspondence of our ACE potentials with experiments, DFT, and related GAP simulations, we consider the computational performance obtained through parallelization and GPU acceleration, and how this compares to recently published models. As already stated, we obtain roughly 1000 times speedup compared to GAP when run on the same resources.  In addition, ACE is already coded to take advantage of GPU parallelization for further acceleration. Figure~\ref{fig:GPU}(a) shows the number of hours per nanosecond of MD simulation using \indirEXP\ with varying resources. All simulations considered use a 1~femtosecond (fs) timestep with system sizes up to 1 million atoms. Though exact simulation times are hardware-dependent these data explicitly demonstrate the approximate speed and scaling of this ACE ML model. We include the case of 1 CPU core as a reference, as well as 8 CPU cores, which is the resource allocation we choose for most simulations considered in previous sections. We also include simulations with GPU acceleration -- for the cases of 1, 2, and 4 GPU's -- which shows a speedup of up to an additional order of magnitude. From these data, it is apparent that GPU acceleration allows for nanosecond device-scale simulations (100,000 - 1,000,000 atoms) to be run on the order of days using a single computational node. We demonstrate this capability in the next section. To more clearly emphasize the relative speedup obtained with GPU simulations, Figure~\ref{fig:GPU}(b) plots the acceleration achieved by the ACE model for various GPU configurations relative to the single CPU core case. Using 4 GPUs and 4 cores, we are able to achieve speeds about 76.5 times faster than the one CPU core case, or about 10.2 times faster than when using 8 CPU cores. This allows us to increase the simulation scale by an additional order of magnitude while maintaining realistic compute times.  These data also indicate that the benefits of GPU acceleration are fully realized for $\approx 1,000$ atoms on a single CPU/GPU setup, and $\approx 10,000$ for a 4 CPU/GPU setup.  Note that, at least for the hardware used here, we found matching the number of CPU cores and GPUs yielded the optimal balance of performance and resource usage.   

We compare our performance findings with several recent studies.  The 2023 GAP model~\cite{zhou2023device} reports 532,980 atoms for a duration of 50 ps with a timestep of 2~fs, or $\approx 1.3\times 10^{10}$ particle updates.  For comparison, the device scale simulations presented in the next section have over 80 times as many particle updates. Earlier this year, the NN potential of El Kheir \textit{et al.}~\cite{kheir2024unraveling}  reports simulations of $\approx$ 12000 atoms for 100 ns, which which is about half the number of updates used in our device scale simulations.  El Kheir \textit{et al.}~\cite{kheir2024unraveling} do not specify what resources were needed for their calculations, making it difficult to  compare performance quantitatively. That said, our larger simulations took modest amounts of time on local resources, and could easily be scaled to even larger and longer simulations. Hence, the ACE potential appears highly competitive from a performance perspective.

\subsection{\label{sec:Device-Scale}Device-Scale Simulations}

\begin{figure*}
\centering
\includegraphics[width=1.0\textwidth,keepaspectratio]{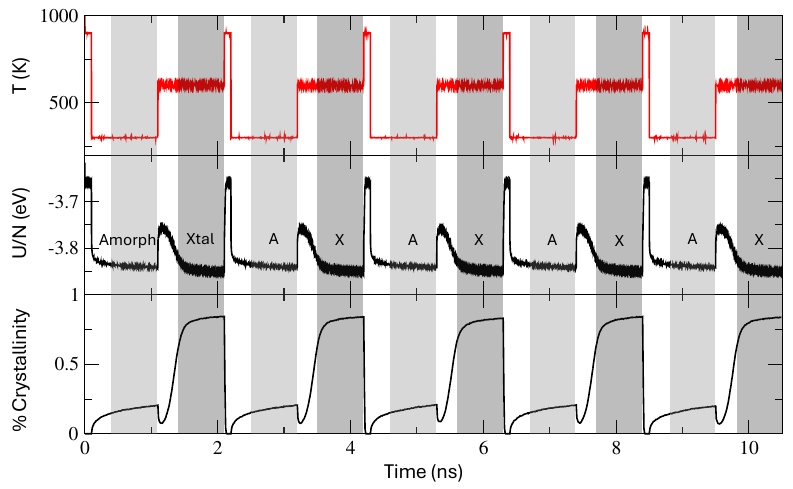}
\caption{A demonstration of the simulation scale readily accessible using the ACE model: multiple phase transitions associated with periodic thermal cycling over 10.5 ns of simulation of a device scale sample of GST (100,800 atoms). The top plot shows the temperature, which we manually cycle between $T=900$ (melt), $T=300$~K (quench to glass), and $T=600$~K (crystallize). The middle and bottom panels show the corresponding change in phase. The middle plot demonstrates the relaxation of the potential energy as the system equilibrates in the amorphous and crystalline phases, while the bottom plot shows the percent of the system identified as crystalline. The light and dark grey shading correspond to the amorphous and crystalline steps, respectively.}
\label{fig:cycle}
\end{figure*}

Due to the computational efficiency of the base ACE model, combined with GPU acceleration, we can readily perform device-scale GST simulations. We use \indirEXP\ to simulate the spontaneous crystallization of a  1,008,000 atom system over 1~ns. This 1 million atom system is significantly larger than those considered in previous works~\cite{kheir2024unraveling, choi2024study, zhou2023device}.  The crystallization simulation was completed using 4 GPUs and 4 CPU cores in just over 10 days. The GST system begins at a 900~K liquid state and is immediately quenched to 600~K where it is held for the duration of the simulation. Figure~\ref{fig:million} shows the temperature, potential energy per atom, and the fraction of atoms identified as crystalline throughout the 1~ns simulation. To identify crystalline atoms, we use the common method developed by ten Wolde \textit{et al.}~\cite{wolde1996numerical}, which we describe in the supplementary information. While annealing at 600~K,  crystallization is indicated by the drop in potential energy and corresponding growth in percent crystallinity.  The plateau in these quantities for $t\gtrsim 600$~ps  suggests  a nearly complete phase transition.  Note that the percent crystallinity never reaches one due to the presence of naturally occurring defects in the crystal structure, as well as misaligned crystal grains.

With the ability to simulate large system sizes and timescales, our ACE potential can readily model repeated transitions between the crystal and amorphous phases in a device-scale system, mimicking the transitions utilized to encode binary states. As a proof of principle, we use \indirEXP\ to simulate multiple phase changes by subjecting a 100,800-atom system to periodic thermal cycling. This simulation is done in the isothermal-isobaric ensemble (NPT), where the pressure is fixed at 1 bar to allow the density to change with the temperature and phase, though very similar behavior occurs if the density (rather than pressure) is fixed.  As we cycle temperature, the phase of the system can be identified by tracking the potential energy and crystallinity (figure~\ref{fig:cycle}). The  thermal cycle can be described in three steps. (i) To mimic a rapid heating pulse, the system is held at 900~K for 100~ps, allowing it to fully melt. (ii) We then quench the system to 300~K, below the glass transition temperature; the resulting amorphous solid anneals at 300~K for 1~ns to confirm the mechanical stability of the amorphous phase. The quench to the amorphous phase is characterized by a sharp decline in the potential energy. There is also a slight increase is the percent of crystallinity, as a fraction of atoms exhibit enough local order to be characterized as crystalline even in the amorphous phase, though none of these nucleate a complete crystallization.  (iii) We then heat the system to 600~K for 1~ns where GST crystallizes over about 500~ps, as shown by the growth in crystallinity towards a predominantly crystalline system (though defects preclude the fraction of crystal atoms from reaching unity). We then repeat this process by reheating the crystal to 900~K and following the above steps. Over the 10.5~ns simulation, this liquid-glass-crystal switching cycle occurs 5 times. 

This procedure is a simplified but relevant reproduction of how the transition between crystal and amorphous GST occurs on-chip to encode binary data. It also demonstrates that our trained potentials can facilitate several complete transitions between ordered and disordered phases on readily accessible timescales. By simply modifying the specifications of the MD simulation, this potential can serve as a tool to study a host of properties of GST and its phase transition.

\subsection{\label{sec:Limitations}Limitations of the Potential}

\begin{figure*}
\centering
\includegraphics[width=1.0\textwidth,keepaspectratio]{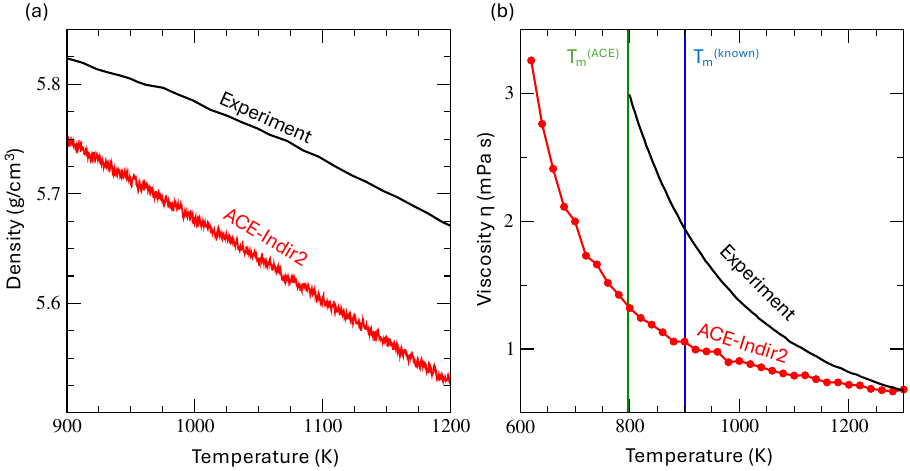}
\caption{Stress-dependent quantities where \indirEXP\ deviates from  experimental~\cite{giessen1972density, schumacher2016viscosity} behavior: (a) density as a function of temperature of liquid GST at atmospheric pressure and (b) viscosity as a function of temperature of liquid GST at experimental density, as well as the melting temperature.  Note that the density deviations are less than $2.5$\%.}
\label{fig:Problems}
\end{figure*}
\justifying

The accurate reproduction of structural properties of GST and the computational efficiency are strengths of our ACE potentials. On the other hand, it is important to recognize that there are limitations on the degree to which the ACE model (or any model) can quantitatively reproduce all experimental properties. It is well known that the pressure is very sensitive to the details of the intermolecular potential, and so the pressure scale of molecular models often does not fully align with experiments. We consider this possibility by examining the density at atmospheric conditions. Specifically, Figure~\ref{fig:Problems}(a) shows the liquid density as a function of temperature during cooling of GST at $P=1$ bar at a rate of 20~K/ns, slow enough to allow sufficient time for the density to equilibrate over the plotted temperature range. The ACE model systematically underestimates the density as compared to experiments~\cite{giessen1972density}. Stated another way, if we were to simulate at a given density, the pressure would be overestimated as compared to the experiment. That said, note that the difference from the experimental density ranges from just $\approx 1\%$ to $2.5\%$ over the entire temperature range, so the error is modest. Given that reproducing the experimental pressure-density relation is a challenge, one can circumvent the issue by manually following a density path that matches the known experimental density.

As another test of the potential, we investigate how well the model reproduces the equilibrium melting temperature $T_m$. This is distinct from the temperature at which crystallization occurs on cooling (or melting occurs on heating) due to metastability.  To evaluate the equilibrium melting temperature, we use the ``interface method''~\cite{vega2008determination} where we consider a crystal and fluid in contact and track the evolution on the interface; if $T<T_m$, the crystal grows, and vice versa above $T_m$. See the SI for details. Using this method we predict the melting temperature $T_m^{(ACE)}=796$~K, which is  smaller than the experimental melting temperature $T_m^{(known)} \approx 900$~K~\cite{bordas1986phase, schumacher2016viscosity, muneer2018activation, kheir2024unraveling}; the precise experimental value differs among published results by a modest amount. For comparison to the two recently published NN potentials, ref.~\citenum{choi2024study} estimates $T_m\approx 1000$~K  and ref.~\citenum{kheir2024unraveling} estimates $T_m \approx 940K$.  These are similar discrepancies to our result, though larger rather than smaller than the experimental value. In the SI, we also consider how including dispersion interactions affects $T_m$; the result is an increase to $T_m=865$~K, considerably closer to the experimental value. 

Matching the dynamical properties of molecular models with the correct experimental values is particularly challenging.  For example, among the multitude of models for water, many exhibit structural properties very similar to experiments but often show substantial differences in their dynamics. 
Thus, as an even more sensitive test of the limitations of the ACE model, we compare the dynamics in the liquid phase of our ML potential to experiments by calculating the shear viscosity $\eta$. We evaluate $\eta$ using the Green-Kubo relation
\begin{equation}
  \eta =  \frac{V}{3k_{B}T} \int_{0}^{\infty} \sum_{\langle ij \rangle} \langle \tau_{ij}(0) \tau_{ij}(t) \rangle dt
\end{equation}
where $\tau_{xy}$, $\tau_{xz}$, and $\tau_{yz}$ are the off-diagonal components of the stress tensor and $\langle \tau_{ij}(0) \tau_{ij}(t) \rangle$ is the stress autocorrelation function. Further details on the calculation of viscosity and convergence of the stress autocorrelation function are available in the supplementary information.
We calculate viscosity at 20~K intervals between 620~K and 1300~K, at densities that are chosen to follow the experimental density path at atmospheric pressure. That said, the experimentally known density range is limited to between 900~K and 1200~K, so we extrapolate the experimental density curve to higher and lower temperatures based on a quartic fit. These calculations are repeated along the natural $P=1$~bar density path of the \indirEXP\ potential, which yields very similar results for viscosity, indicating that the accuracy of the density extrapolation does not significantly affect the calculated viscosity. For each $(\rho, T)$ state point, the system is relaxed for 100~ps, and $\eta$ is then calculated from a 1~ns equilibrium trajectory. 

Figure~\ref{fig:Problems}(b) shows the resulting dependence of $\eta$ from the ACE model in comparison to the experimentally known viscosity~\cite{schumacher2016viscosity}. The viscosity computed by \indirEXP\ is consistently lower than that of the experiment.  This lack of agreement was foreshadowed by the fact that the density at $1$~bar was not consistent with experiments, indicating that the stress tensor is not accurate, and hence viscosity should also be affected. Similar to the underestimation of $\eta$, the diffusion coefficient  is larger than that predicted by DFT (see SI fig.~S6).  
%, and the sharp increase in viscosity towards the glass transition is roughly 150~K too low. 
Thus, if using the \indirEXP\ to examine  the dynamics of the liquid phase, one needs to be cognizant of the fact that the molecular mobility is larger than expected from experiments. The enhanced fluidity of our model may also relate to the fact that the melting temperature is lower than that observed experimentally. Earlier \textit{ab initio} MD simulations using a different functional indicated a significant over-estimate of viscosity compared to experiments~\cite{schumacher2016viscosity}, indicating that viscosity and other dynamic properties may be very sensitive to the choice of functional.

\begin{figure}
\centering
\includegraphics[width=0.4\textwidth,keepaspectratio]{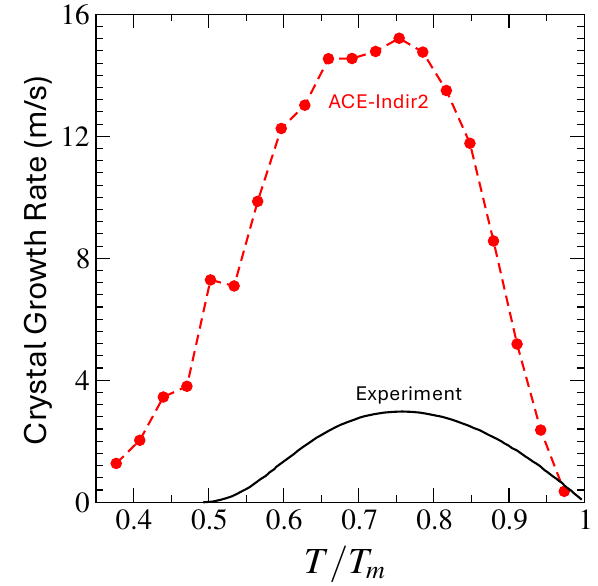}
\caption{Crystal growth rate as a function of temperature as predicted by our ACE model, in comparison with experiment~\cite{orava2012characterization}. The ACE model predicts a larger growth rate than experiment, consistent with the differences in viscosity.  The ACE model does capture the non-monotonic variation of the crystal growth rate. }
\label{fig:growth}
\end{figure}
\justifying

As a result of the increased molecular mobility, we expect the crystallization kinetics to be faster than those measured by experiments, as particles with greater mobility can explore new arrangements at higher rates, and thus reach stable locally crystalline environments more quickly. 
In fact, we find that this is the case for our ACE model. By again using the interface method~\cite{vega2008determination} and tracking the position  of the crystal-liquid interface over time, we determine the mean crystallization rate for \indirEXP\ (see the SI for more explicit details on the calculation). Figure~\ref{fig:growth} shows the crystallization rate as a function of temperature for the ACE model compared to the experimentally measured values~\cite{orava2012characterization}. Importantly, \indirEXP\ captures the non-monotonic temperature dependence of the crystal growth rate found experimentally, with both showing a maximum in the interfacial speed around $T/T_m=0.75$. As expected from $\eta$ and diffusivity, the ACE model shows a growth rate consistently several times faster than experiments. While this is a limitation on the model's ability to quantitatively reproduce crystal kinetics, one can expect that relative behavior of the crystallization speed due to changing external conditions should be predictive of the experimental behavior. Additionally, it should be noted that the ML potential of el Kheir \textit{et al.}~\cite{kheir2024unraveling} better matches the crystallization kinetics.  Since their DFT training set is available, it would  potentially be valuable to test how an ACE potential trained on those data performs.
On the other hand, the accelerated crystallization of this ACE potential enables the rapid examination of the variation of crystallization and structure over a broader range of thermodynamic or confinement conditions.

A possible reason that the model overestimates molecular mobility is that dispersion interactions may need to be explicitly considered.  We add a dispersion correction to the ML potential using the commonly employed D2 dispersion methods of Grimme {\it et al.}~\cite{grimme2006semiempirical, grimme2016dispersion}. However, the DFT data on which our potential energy surfaces are indirectly trained are computed using the PBEsol functional~\cite{perdew2008restoring}; unfortunately, there are no definitive values for the D2 dispersion parameters for this functional. A range of possible values -- including no dispersion correction at all -- have been suggested for PBEsol, depending on the system studied~\cite{ibarra2018ab, fischer2017accurate, dispersion2018terentjev, rybakov2017CO, agbaoye2017elastic, rios2018effects, mairesse2024simple, feigelson2015growth, gu2017high, goerigk2011efficient}. That said, we select D2 parameters using an empirical approach, and evaluate the inclusion of dispersion interactions with the \indirEXP\ potential (see the supplementary information), though these efforts are not successful in improving the viscosity's agreement with experiment.

Apparently, our model is most sensitive to material properties that depend heavily on the stress tensor, such as pressure and viscosity.  Since the ACE model reproduces the DFT per-atom forces that determine the stress tensor, this is likely an indication of the limitations of the DFT model used for training, rather than the ACE potential.  The substantial variation in viscosity predictions between our work and Schumacher \textit{et al.}\ who use a different functional is another indicator of the high degree of sensitivity of stress-tensor dependent quantities on the details of the potential.

\section{Conclusion}

We have used the indirect learning approach to fit an ML interatomic ACE potential which faithfully models the structure, thermodynamic properties, and phase transition of GST, with accuracy comparable to DFT simulations. We have demonstrated the ability of our trained ACE potentials to reproduce the behavior of alternative models -- particularly GAP -- with a computational speedup of 3 orders of magnitude for CPU-only calculations, and an additional 1-2 orders of speedup available via GPU acceleration. We demonstrate the ability of the ACE potentials presented here to simulate a device-scale sample of GST for multiple nanoseconds within reasonable computing time; a similar simulation with the GAP potential or DFT would be unrealistic without employing massive computational resources. Finally, we highlight areas where caution must be taken when making quantitative comparisons with experiments, particularly for material properties that depend on the stress tensor. Given that the ACE model reproduces the DFT forces on which it was trained to a high degree of accuracy, this is probably an indication of the limitations of the DFT calculations.

This procedure serves as validation of the indirect learning approach: we demonstrate physical agreement between the directly learned and indirectly learned potentials, both trained on the GAP training set. Thus, in this case, quantum-mechanical accuracy is not lost when training is done through an intermediate ML potential. Further research is needed to determine the size and diversity of a training set necessary to reach convergent results using the ACE model. This may help to identify scenarios in which generating an arbitrarily large dataset via indirect learning would significantly improve the performance of a trained potential compared to direct learning on a limited and computationally expensive DFT training set.

\begin{acknowledgments}
We thank Carlos Jimenez Hoyos and Jack Douglas for their helpful discussions. 
Computer time was provided by Wesleyan University. This
work was supported in part by NIST awards 70NANB19H137 and 70NANB24H025.
\end{acknowledgments}

\section*{Data Availability Statement}

The trained ACE potentials discussed in this study are openly available via Zenodo at \href{https://doi.org/10.5281/zenodo.12173540}{DOI: 10.5281/zenodo.12173540}. The MD trajectories that support our findings, as well as the configurations used to expand the training set, are available from the corresponding author upon reasonable request.

%\nocite{*}
%\bibliography{main}% Produces the bibliography via BibTeX.

%
% ****** End of file aipsamp.tex ******%merlin.mbs aipnum4-1.bst 2010-07-25 4.21a (PWD, AO, DPC) hacked
%Control: key (0)
%Control: author (8) initials jnrlst
%Control: editor formatted (1) identically to author
%Control: production of article title (-1) disabled
%Control: page (0) single
%Control: year (1) truncated
%Control: production of eprint (0) enabled
%
\end{document}

\title{Supplementary Information for \\ Computationally Efficient Machine-Learned Model for GST \\ Phase Change Materials via Direct and Indirect Learning}
% \author{O. Dunton}
% \author{T. Arbaugh}
 \author{Owen R.\ Dunton, Tom Arbaugh, and Francis W.\ Starr}
\date{November 5, 2024}
\maketitle

\tableofcontents
\newpage

\section{Identification of Local Crystalline Environments for Computation of Percent Crystallinity}

\subsection{Cutoff Radius}

\begin{figure}[!htb]
\centering
\includegraphics[width=1.0\textwidth,keepaspectratio]{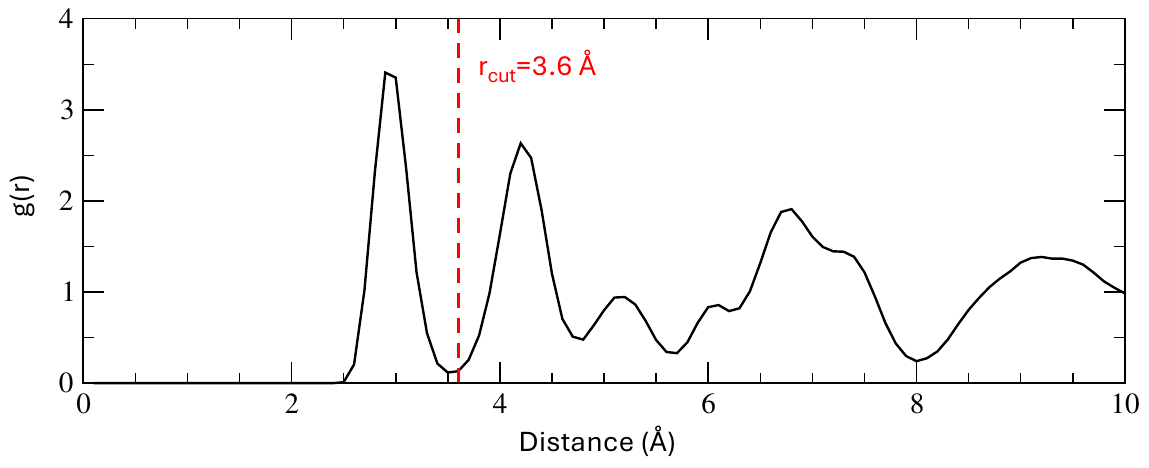}
\caption{Radial distribution function of a spontaneously nucleated bulk GST crystal, with the location of the first minimum labeled as the cutoff distance of an atom's local neighborhood $r_{cut}=3.6$~\AA.}
\label{fig:gr}
\end{figure}

In order to distinguish crystal nuclei from an amorphous background, we employ a standard method discussed by ref.~\cite{wolde1996numerical} to identify which particles form crystal-like arrangements with nearest neighbors. We define the neighborhood of an atom $i$ by building a neighbor list of all $N_{n}(i)$ atoms $j$ such that $|\mathbf{r}_{ij}|<r_{cut}$, where $\mathbf{r}_{ij}$ is the vector to atom $j$ from atom $i$, and $r_{cut}$ is a cutoff distance. We determine this cutoff distance by considering the radial distribution function $g(r)$ for a spontaneously nucleated GST crystal (Fig.~S\ref{fig:gr}), where the first peak location represents the average distance of a given atom's nearest neighbors. The location of the first minimum $r_{cut}=3.6$~\AA\ is selected, such that only the first ring of neighbors is considered. 

\justifying

\subsection{Description of Local Structure}

To describe the structure of the local environment of atom $i$, its neighbors are projected onto spherical harmonics $Y_{lm}$ of a particular index $l$~\cite{wolde1996numerical}. These $2l+1$ projections are given by 
\begin{equation}
    \bar{q}_{lm}(i)=\frac{1}{N_n(i)} \sum_{j=1}^{N_n(i)}Y_{lm}(\mathbf{\hat{r}}_{ij})
\end{equation}
where $\mathbf{\hat{r}}_{ij}$ is a unit vector and $-l \leq m \leq l$. The $l$-index is selected based on the lattice structure of the crystal considered, as different choices for $l$ may provide different signal strengths, depending on the crystal geometry. Specifically, $l=6$ is optimal in the cases of fcc and bcc lattices, while $l=4$ provides a stronger signal in the case of a simple cubic (sc) lattice. The lattice structure can be determined using the crystalline order parameters 
\begin{equation}
    Q_l=\sqrt{\frac{4\pi}{2l+1} \sum_{m=-l}^l |\bar{Q}_{lm}|^2}
\end{equation}
where $\bar{Q}_{lm}= \left[ \sum_{i=1}^N N_b(i) \bar{q}_{lm}(i) \right] / \left[ \sum_{i=1}^N N_b(i) \right]$. Table~S\ref{table:potentials} gives the expected order parameters $Q_4$ and $Q_6$ for a perfect simple cubic lattice~\cite{wolde1996numerical} as well for the lattice structure obtained in our simulations via spontaneous nucleation. The order parameters for the simulated crystal are slightly lower than for the case of a perfect crystal, due in part to stochastic imperfections that emerge in crystal nucleation, as well as to inherent vacancies in GST's crystal structure resulting from its stoichiometry. Despite this, these order parameters affirm that our simulated GST has a simple cubic crystal structure. Accordingly, $\bar{q}_{4m}(i)$ provides the strongest signal of local crystalline structure in GST.

\begin{table}[!htb]
\caption{Crystalline order parameters for a sc lattice\cite{wolde1996numerical} and a spontaneously nucleated GST crystal in our simulations.}
\centering
\begin{tabular}{ |>{\centering\arraybackslash}p{2.6cm}||>{\centering\arraybackslash}p{1.5cm}|>{\centering\arraybackslash}p{1.5cm}| }
 \hline
 %\multicolumn{4}{|c|}{Potentials Discussed} \\
 %\hline
   & $Q_4$ & $Q_6$\\
 \hline
SC Lattice & 0.764 & 0.354 \\
GST Simulation & 0.71669 & 0.31805 \\
 \hline
\end{tabular}
\label{table:potentials}
\end{table}

\justifying

For every atom $i$, a 9-dimensional unit vector is built with an entry corresponding to the $m-$indices from $-4$ to $4$. This vector is given by
\begin{equation}
    \mathbf{q}_{4}(i) = \frac{1}{\sqrt{\sum_{m=-4}^4 |\bar{q}_{4m}(i)|^2}}
    \begin{pmatrix}
    \bar{q}_{4,4}(i) \\
    \bar{q}_{4,3}(i) \\
    \vdots \\
    \bar{q}_{4,-4}(i) \\
    \end{pmatrix}.
\end{equation}
For each atom $i$, the dot product $\mathbf{q}_{4}(i) \cdot \mathbf{q}_{4}(j)$ is taken for each of its $N_n(i)$ neighbors $j$, and neighbors $i$ and $j$ are identified as having ``connected'' crystalline environments if the real part of this dot product exceeds a cutoff value $\Re(\mathbf{q}_{4}(i) \cdot \mathbf{q}_{4}(j) ) > \xi$. While there is nothing preventing a strong alignment between two atom environments from occurring randomly in an amorphous system, agreement over many neighbors is unique to local regions of crystalline order. Thus, a particle is deemed crystalline if it has more than $\text{n}_{\text{xtal}}$ connected crystalline neighbors. 

\subsection{Optimization of $\xi$ and $\text{n}_{\text{xtal}}$}

\begin{figure}[!htb]
\centering
\includegraphics[width=1.0\textwidth,keepaspectratio]{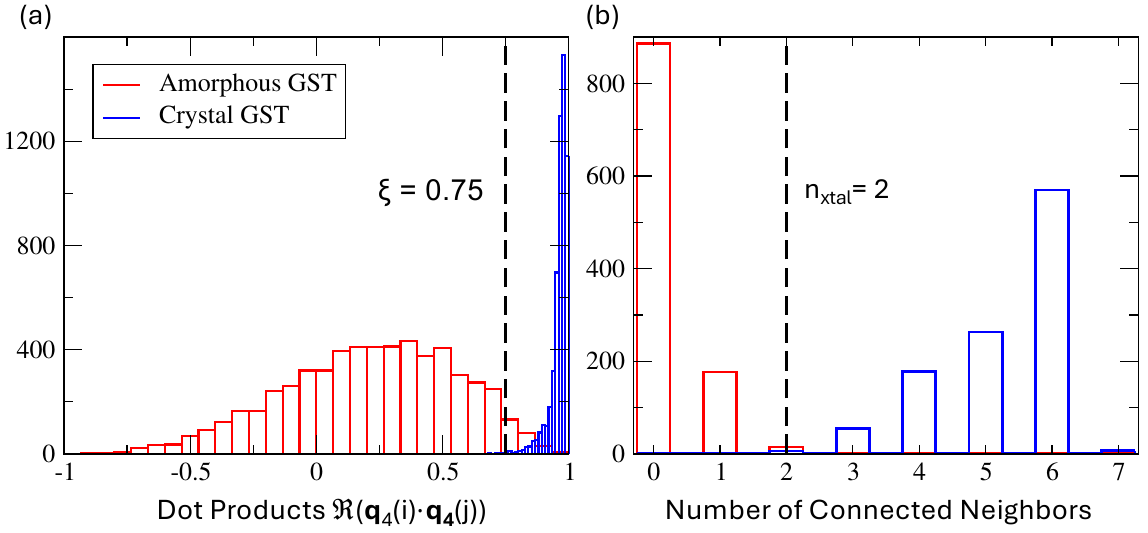}
\caption{Plots used to optimize the parameters $\xi$ and $\text{n}_{\text{xtal}}$ using liquid GST equilibrated at $T=900$~K and crystalline GST equilibrated at $T=600$~K. (a) A histogram of the real parts of all dot products $\Re(\mathbf{q}_{4}(i) \cdot \mathbf{q}_{4}(j) )$ between neighboring atoms of the liquid and crystal systems. (b) A histogram of the number of connected crystalline neighbors predicted for crystal and amorphous GST when $\xi=0.75$ is selected.}
\label{fig:parametes}
\end{figure}

The values $\xi$ and $\text{n}_{\text{xtal}}$ are selected to best distinguish GST's crystalline and amorphous structures. We optimize these parameters using a crystalline GST system spontaneously nucleated at $T=600$~K and a liquid GST system equilibrated at $T=900$~K, both of 1080 atoms. To determine $\xi$, we create a histogram all dot product values $\Re(\mathbf{q}_{4}(i) \cdot \mathbf{q}_{4}(j) )$ between neighboring atoms, shown in Fig.~S\ref{fig:parametes}(a) for both crystalline and amorphous GST. From this, we select $\xi=0.75$, since nearly all dot products between crystalline GST neighbors lie above this value, while only a small number of neighbors in an amorphous environment exhibit $\Re(\mathbf{q}_{4}(i) \cdot \mathbf{q}_{4}(j) ) > 0.75$. With $\xi=0.75$ selected, Fig.~S\ref{fig:parametes}(b) shows a histogram of all atoms in the crystalline and amorphous systems as a function of the number of connected neighbors. From this, we select $\text{n}_{\text{xtal}}=2$, as nearly all amorphous atoms have 2 or fewer connected crystalline neighbors, while the vast majority of crystalline atoms have more than 2 connected neighbors.

As a quantitative validation of this parameter choice ($r_{cut}=3.6$~\AA, $\xi=0.75$, $\text{n}_{\text{xtal}}=2$), we  identify crystalline particles in the bulk crystal system, where all atoms are expected to have crystalline local environments, and the bulk liquid system, where there should be no crystalline particles. In the crystal system, the model identifies 1072 out of 1080 atoms as crystalline, suggesting that the model fit with these parameters accurately identifies crystalline particles with a probability of 99.35\%; the misidentified particles are a results of thermal defects.  Likewise, only one particle in the amorphous system is predicted to have a crystal local environment, suggesting an error rate of 0.09\%. 

These results show that this method nearly perfectly classifies the structure of bulk GST. We apply the method for the identification of growing crystal nuclei in a fluid; since these simulations must be done below the crystallization temperature, it is not uncommon for particles to transiently exhibit crystalline geometries without forming a nucleus large enough to grow, resulting modest background of isolated crystalline particles. This makes it difficult to distinguish crystal nuclei from the amorphous background. To eliminate this background, we impose the additional constraint that lone particles cannot be considered crystalline. Once all atoms with more than $\text{n}_{\text{xtal}}=2$ connected neighbors are identified, we impose this constraint by removing any atoms without at least one neighboring crystalline particle. 

\section{\label{Interface}Interface Method for Melting Temperature \& Crystal Growth Rate}

\begin{figure}[!htb]
\centering
\includegraphics[width=1.0\textwidth,keepaspectratio]{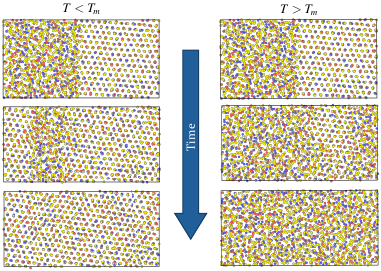}
\caption{Snapshots of the same initial configuration as it evolved below and above the melting temperature, with the equilibrium phase growing as the crystal-amorphous interface evolves.}
\label{fig:melting}
\end{figure}

The melting temperature cannot be reliably determined through simple heating or cooling simulations due to metastability of the phases, which often leads to strong hysteresis effects.   Thus, we use the interface method~\cite{vega2008determination}, where we start with an unbiased initial configuration (half-crystalline and half-liquid) with a defined crystal-liquid interface; depending on the temperature, either the liquid or crystal will grow.  These simulations are done with 2160 atoms where the cell dimensions are fixed in the $\hat{x}$ and $\hat{y}$ directions (parallel to the interface); in the $\hat{z}$ direction, we maintain a constant $P_{zz}=1$~bar which allows the box size to change normal to the interface and thus the density changes as needed due to melting or crystallization. As shown in fig.~S\ref{fig:melting}, the crystal-liquid interface will evolve in favor of the equilibrium phase. By performing these simulations at resolutions of 1~K near the melting temperature, we can predict its value within $\pm 2-3$~K (within 2-3~K of the melting temperature,  thermal fluctuations dominate the evolution of the interface, as opposed to the equilibrium phase, on the timescales accessible via MD). Specifically, we converge to a melting temperature of $T_{m}^{(ACE)}=796$~K $\pm 2$~K; the experimental melting temperature is $T_m^{(known)}=903.15$~K~\cite{bordas1986phase}.

Similarly, this method can determine the crystal growth rate at a chosen temperature below $T_m$. First, we perform a crystallization simulation, as described above, from a half-crystalline and half-liquid initial configuration. By discretizing the $\hat{z}$-axis, we can histogram the percent crystallinity of the system along the $\hat{z}$ direction, where the interface becomes apparent as a sharp increase from the majority of particles being identified as amorphous to the majority being identified as crystalline. We can track the evolution of this interface in time to extract the crystal growth rate. Due to the stochastic nature of crystal growth, the noise in these data is significant, so the growth rates reported in the main body of the paper are averaged over 8 independent simulations.

\section{Viscosity Calculation Details}

\subsection{\label{sec:density}Density Choice}

\begin{figure}[!htb]
\centering
\includegraphics[width=1.0\textwidth,keepaspectratio]{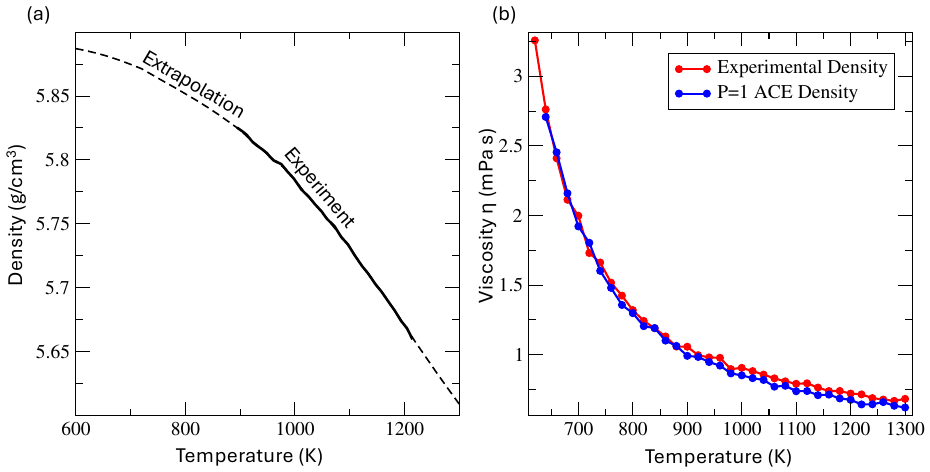}
\caption{(a) The experimental density~\cite{giessen1972density} of liquid GST, with the quartic extrapolation to higher and lower temperatures shown by the dashed line. (b) The viscosity of liquid GST, as predicted by \indirEXP, at both the experimental density and the $P=1$~bar natural density of the ACE potential.}
\label{fig:density_choice}
\end{figure}

We wish to compare our estimates of viscosity with experiments along the experimentally known density path. However, the experimental~\cite{giessen1972density} liquid GST density is limited to temperatures between 900~K and 1200~K, and thus we use the quartic extrapolation shown in Fig.~S\ref{fig:density_choice}(a) to choose densities for viscosity calculations outside of this temperature range. While the validity of this extrapolation is not known, we shall see that that small changes of density do not significantly affect the viscosity of this ML potential. In fact, we supplement the viscosity calculation at the experimental density with an identical calculation for systems along the $P=1$~bar equilibrium density path for our ACE model. As shown in the main body of the manuscript, the natural $P=1$~bar density of this potential differs from experiment by $\approx 1\%$ to $2.5\%$. That said, Fig.~S\ref{fig:density_choice}(b) shows that viscosities computed along these two density paths are nearly identical. Thus, we conclude that the degree to which the extrapolation may vary from experiment does not significantly impact the reported viscosity.

\subsection{Convergence of the Stress Autocorrelation Function}

\begin{figure}[!htb]
\centering
\includegraphics[width=1.0\textwidth,keepaspectratio]{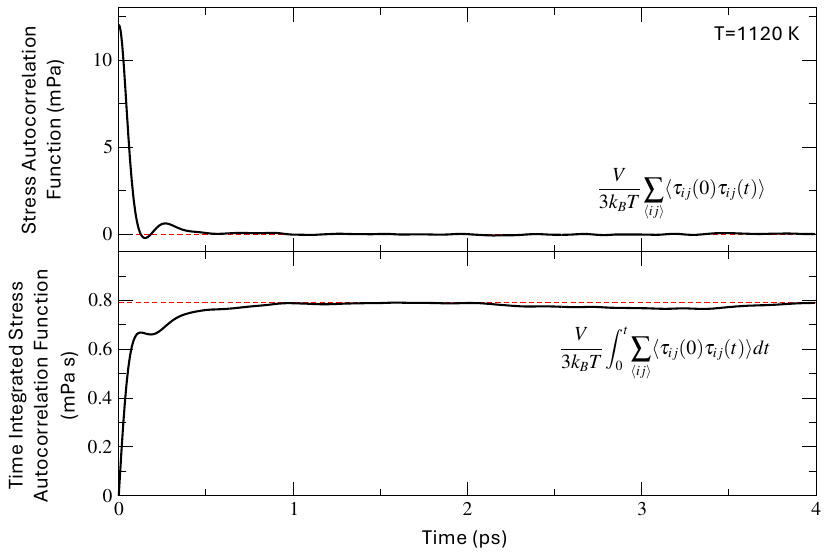}
\caption{The autocorrelation function of the off-diagonal components of the stress tensor, and its time integral, used to calculate viscosity.}
\label{fig:autocorrelation}
\end{figure}

While the analytic calculation of the viscosity using the Green-Kubo method requires the autocorrelation function off-diagonal elements of the stress tensor to be integrated over infinite time $\int_{0}^{\infty} \langle \tau_{ij}(0) \tau_{ij}(t) \rangle dt$, it is sufficient to integrate over a finite simulation $\int_{0}^{t'} \langle \tau_{ij}(0) \tau_{ij}(t) \rangle dt$ so long as the autocorrelation function is well converged to 0 by $t=t'$. 

These simulations are performed at constant temperature every 20~K between 600~K and 1300~K. We run these simulations from temperature-dependent initial configurations prepared at the density predicted in sec.~\ref{sec:density}, and equilibrated for 200~ps at the selected temperature. We then calculate the viscosity from 5~ns simulations in the NVT ensemble, where the autocorrelation function of stress tensor components $\tau_{xy}$, $\tau_{xz}$, and $\tau_{yz}$ are independently calculated using the LAMMPS \verb|fix ave/correlate| command %. The autocorrelation functions have a correlation length of $t'=4$~ps, 
with a resolution of 5~fs. By performing this calculation over 5 ns, the autocorrelation value at each time $\{ \langle \tau_{ij}(t_1) \tau_{ij}(t_2) \rangle; \ t_2-t_1=5~fs,\ 10~fs,\ ...\ 4~ps \}$ is calculated about 1 million times (though those at large time differences are calculated slightly fewer). Thus, the stress autocorrelation functions used to calculate the viscosity are averaged over $\approx 1,000,000$ intervals during the simulation. We then average the autocorrelation functions over the off-diagonal components of the stress tensor 
\begin{equation}
    \frac{V}{3k_{B}T} \left[ \langle \tau_{xy}(0) \tau_{xy}(t) \rangle + \langle \tau_{xz}(0) \tau_{xz}(t) \rangle + \langle \tau_{yz}(0) \tau_{yz}(t) \rangle \right],
\end{equation}
which gives the stress autocorrelation function used to predict the viscosity for a given simulation. The plot of this function for an example simulation at $T=1120$~K is shown in the top plot of Fig.~S\ref{fig:autocorrelation}, where it is clear that the function is converged to 0 within a $4$~ps correlation window. We then calculate the viscosity by integrating this function over this correlation length 
\begin{equation}
    \eta = \frac{V}{3k_{B}T} \int_{0}^{t'} \left[ \langle \tau_{xy}(0) \tau_{xy}(t) \rangle + \langle \tau_{xz}(0) \tau_{xz}(t) \rangle + \langle \tau_{yz}(0) \tau_{yz}(t) \rangle \right]dt.
\end{equation}
To demonstrate that this integral is also well converged, we plot its running value over the correlation length in the bottom plot of Fig.~S\ref{fig:autocorrelation} with the final predicted viscosity shown in red. Though the viscosity is clearly well converged by 4~ps, there is more noise in the integrated autocorrelation function than in the function itself. To minimize the impact of these fluctuations, we report the viscosity at each temperature averaged over 5 independent  5~ns simulations.

\section{Diffusion Coefficient}

\begin{figure}[h]
\centering
\includegraphics[width=0.5\textwidth,keepaspectratio]{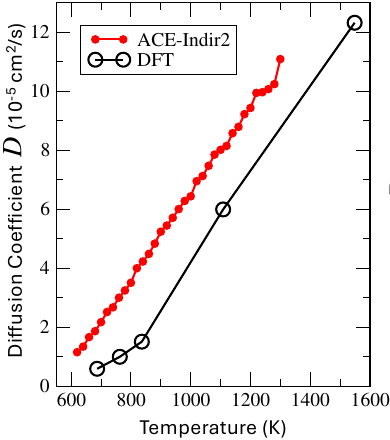}
\caption{The GST self-diffusion coefficient as a function of temperature, as predicted by \indirEXP, compared with published DFT data~\cite{rizzi2014statistics}.}
\label{fig:DSE}
\end{figure}

As an additional measure of mobility in our GST ACE model, we evaluate the self-diffusion coefficient of liquid GST as a function of temperature. We compute $D$ as the slope of the diffusive regime of the mean-squared displacement and plot the results in Fig.~S\ref{fig:DSE}. These are plotted along with published DFT calculations of the diffusion coefficient~\cite{rizzi2014statistics}, showing that the ACE model  predicts a diffusion coefficient that is higher than the known values, as expected from the fact that the viscosity is underestimated compared to experiments. 
%It is worth noting however that these two models predict the diffusion coefficient with comparable linear behavior and slope with respect to temperature. 

Additionally, using the diffusion coefficients plotted here and the viscosity $\eta$ reported in the main body, we calculate Stokes-Einstein ratio $D\eta / T$ and find that it is constant within the noise of the data, and thus the Stokes-Einstein relation is valid over this limited temperature range.  Typically, the Stokes-Einstein relation fails nearing the glass transition.

\section{D2 Dispersion Correction}

\begin{figure}[!hb]
\centering
\includegraphics[width=1.0\textwidth,keepaspectratio]{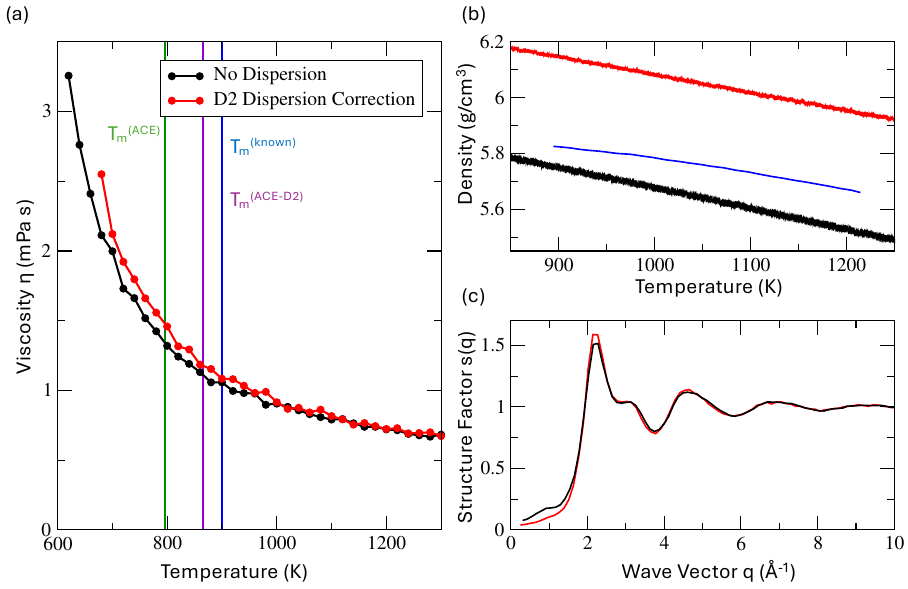}
\caption{Various GST properties predicted by \indirEXP\ with and without D2 dispersion corrections, including (a) the viscosity and melting temperature (b) the density and (c) the liquid structure factor.}
\label{fig:dispersion}
\end{figure}

Generalized gradient approximations, including the PBESol functional, are  known to underestimate the effects of dispersion. We recognize that the omission of explicit dispersion corrections in the potential can alter the dynamics of our GST simulation. The inclusion of dispersion corrections may increase viscosity, which would bring our results more in line with experiments. Hence, we consider how using  the D2 correction scheme -- originally proposed by Grimme~\cite{grimme2006semiempirical} and widely employed in DFT studies -- affects our results.  In this scheme, the energy (and thus forces) are corrected by a factor
\begin{equation}
    E_i^{D2} = -s_6 \sum_{j \neq i}^N (\frac{C_6^{ij}}{R_{ij}^6}) f_{damp}(R_{ij})
\end{equation}
where $C_6^{ij}$ is the dispersion coefficient for the pair $ij$ and $s_6$ is a scaling parameter that depends on the functional used. The damping function
\begin{equation}
    f_{damp}(R_{ij}) =\left(1 + \exp \left( {-20(\frac{R_{ij}}{s_r(r_i+r_j)} - 1)} \right) \right)^{-1}
\end{equation}
where $r_i$ and $r_j$ are the van der Waals radii for elements $i$ and $j$; $s_r$ is a scaling factor that again depends on the functional. Unfortunately, for the the PBEsol functional (used to create the training configurations) there are no definitive values of the scaling parameters $s_6$ and $s_r$; various works recommend values of $s_6$ as small as 0 (no correction) to 1.5, while $s_r$ is typically in the range 1.1 to 1.15, depending on the system studied~\cite{ibarra2018ab, fischer2017accurate, dispersion2018terentjev, rybakov2017CO, agbaoye2017elastic, rios2018effects, mairesse2024simple, feigelson2015growth, gu2017high, goerigk2011efficient}. Thus, we take an empirical approach and select parameters that retain the strong agreement of the structural properties with experimental data, which leads us to choose $s_6=0.75$ and $s_r=1.15$; these values are close to the values originally suggested for the related PBE functional. For reference, Fig.~S\ref{fig:dispersion}(c) shows that the liquid structure factor $S(q)$ of \indirEXP\ corrected by D2 dispersion interactions with these parameters $s_6=0.75$ and $s_r=1.15$ agrees with that of the uncorrected potential. Using larger values of $s_6$ or not scaling the radii ($s_r=1$) significantly alters the first two peaks of $g(r)$ as compared to \indirEXP, and thus, experiments.

Having chosen D2 parameters that preserve the experimentally accurate structure, we can explore the impact of dispersion corrections on the properties where \indirEXP\ diverges from experimental results. Using the interface method, we reevaluate the equilibrium melting temperature and find that the melting temperature of the D2 corrected ACE potential $T_m^{(ACE-D2)}=865$~K is noticeably closer to the experimentally measured value $T_m^{(known)}=903.15$~K~\cite{bordas1986phase}, compared with the uncorrected ACE potential $T_{m}^{(ACE)}=796$~K. 

That said, including dispersion does not significantly alter the viscosity, as shown in Fig.~\ref{fig:dispersion}(a). In fact, the dispersion-corrected ACE potential has nearly the same viscosity as the uncorrected version, with a very slight increase at low temperatures. In addition to this D2 parameter choices -- selected to retain the structure -- we sampled the viscosity for a variety of other parameter choices. Ultimately, some parameter choices increased the viscosity beyond what is shown here, but none of them were nearly sufficient to restore quantitative agreement with the experimentally known viscosity. In contrast, viscosity estimates from direct AIMD simulations using a different functional~\cite{schumacher2016viscosity} actually overestimate viscosity compare with experiments.  We conclude that the increased molecular mobility of our ACE model is not simply a result of omitting dispersion corrections.
%, but instead must be addressed in the training process through the consideration of the stress tensor.

Finally, Fig.~\ref{fig:dispersion}(b) shows the equilibrium density vs. temperature curves at atmospheric pressure as predicted by the ACE model with and without dispersion corrections. The inclusion of dispersion interactions significantly increases the density. While we had hoped for a slight increase in density to make up the $\approx 1\%$ to $2.5\%$ discrepancy between the ACE potential and experiment, the dispersion-corrected model overshoots this goal and predicts a density that is $\approx 5\%$ too high.

\bibliographystyle{aip}
%\bibliography{main}% Produces the bibliography via BibTeX.